\documentclass[aps,nobalancelastpage,amsmath,amssymb,a4paper,twocolumn]{revtex4}

\usepackage{graphics}     
\usepackage{url}          
\usepackage{bm}           
\usepackage[english]{babel}    
\usepackage[latin1]{inputenc}  
\usepackage[dvips]{graphicx} 
\usepackage{mathrsfs}
\def\<{\langle}
\def\>{\rangle}

\def\½{\frac{1}{2}}
\def\klaser{k_{{\text{\tiny L}}}}
\def\laser{\omega_{{\text{\tiny L}}}}
\def\stokes{\omega_{{\text{\tiny S}}}}
\def\kstokes{k_{{\text{\tiny S}}}}

\numberwithin{equation}{section}
 
\begin{document}

\title{Three-dimensional theory of stimulated Raman scattering}
\author{Martin W. Sørensen} \affiliation{QUANTOP -- Danish quantum
  optics center and the Niels Bohr Institute, University of
  Copenhagen, DK-2100 Copenhagen \O, Denmark}
\author{Anders S. S\o rensen} \affiliation{QUANTOP -- Danish quantum
  optics center and the Niels Bohr Institute, University of
  Copenhagen, DK-2100 Copenhagen \O, Denmark}

\date{\today}

\begin{abstract}
  We present a three-dimensional theory of stimulated Raman scattering
  (SRS) or superradiance. In particular we address how the spatial and
  temporal properties of the generated SRS beam, or Stokes beam, of
  radiation depends on the spatial properties of the gain medium.
  Maxwell equations for the Stokes field operators and of the atomic
  operators are solved analytically and a correlation function for the
  Stokes field is derived.  In the analysis we identify a
  superradiating part of the Stokes radiation that exhibit beam
  characteristics. We show how the intensity in this beam builds up in
  time and at some point largely dominates the total Stokes radiation
  of the gain medium.  We show how the SRS depends on geometric
  factors such as the Fresnel number and the optical depth, and that
  in fact these two factors are the only factors describing the
  coherent radiation.
\end{abstract}

\maketitle

  \section{Introduction}

  The collective emission of radiation from an ensemble of atoms is
  interesting both from fundamental as well as an applied perspective.
  The enhanced collective emission or superradiance \cite{haroche}
  from an atomic ensemble was predicted already by Dicke in 1954
  \cite{dicke} and was observed in the form stimulated Raman
  scattering (SRS) in 1962 \cite{woodbury}, but in recent years it has
  attracted renewed interest due to the the observation of SRS from
  Bose Einstein condensates Ref.
  \cite{ketterle1,ketterle2,sadler,hilliard08}.  From a more applied
  perspective the problem of SRS is closely related to free
  electron lasers \cite{fel} as well as to activities in quantum
  information science aiming at realizing a quantum interface between
  light and matter \cite{hammerer}. Recently it has even been proposed
  that SRS from a Bose-Einstein condensate could serve as a direct
  source of entanglement \cite{polzik-cirac1}.

  From a theoretical perspective one of the challenges consists of
  describing SRS from an extended ensemble. Whereas the original Dicke
  superradiance \cite{dicke} was described for a collection of atoms
  localized to dimensions much less than the wavelength of the
  outgoing light, most experiments are actually performed in the
  opposite regime where the dimensions of the ensemble is much larger
  than the wavelength. A full quantum description of SRS was presented
  by Raymer and Mostowski in Ref.  \cite{raymer-mostow} using a one
  dimensional model. Such a one-dimensional description can be shown
  to be applicable to all transverse modes of the field if the gain
  medium has an infinite transverse extension \cite{hammerer}. In such
  a description there is, however, no restriction on the transverse
  modes and a summation over all transverse modes therefore results in
  an infinite intensity of the outgoing light. The theory was
  generalized to also include certain three-dimensional properties
  of the propagation of light in the gain medium in Ref.
  \cite{raymer-mostow2}. Here it was argued that the one-dimensional
  theory could be used to predict the total intensity for a sample
  with a Fresnel number of unity $\mathcal{F}=1$, where the process
  was dominated by a single transverse mode.  These theories were
  developed under the basic assumption that the region in which this
  SRS process happens is defined by the properties of the laser both
  in time and space. Thus figures of merits are the width and temporal
  shape of the laser which is driving the SRS process. The experiments
  exploring the SRS process have changed since then
  \cite{ketterle1,ketterle2,sadler,hilliard08}, and much more attention is given
  to systems where the temporal and spatial shape of the laser have
  long surpassed the spatial geometries and temporal properties of the
  gain medium. A three dimensional theory applicable to small atomic
  ensembles were presented in Refs. \cite{moore,zobay} in the
  approximation that certain off-diagonal matrix elements in the
  momentum representation could be ignored. For the closely related
  problem of light emission from an ensemble of atoms with a few
  collective excitations, direct numerical simulations have been
  performed for a few thousand atoms \cite{akkermans,hjortshoj}. Using
  periodic boundary conditions the three dimensional effects of this
  problem was studied in Ref.  \cite{cirac}, and an approximate
  analytical treatment was also presented in Ref. \cite{Svidzinsky}.
  To our knowledge, however, no theory have been developed which fully
  describe SRS from a spatially extended ensemble.  Here we
  develop the theoretical framework that enables us to describe
  SRS from such extended ensembles. Since we shall neglect
  the depletion of the initial atomic state, the present theory is,
  however, only capable of describing the onset and build up of
  SRS. In the resulting theory the only two parameters
  describing superradiance is the optical depth $d$ and the Fresnel
  number of the sample $\mathcal{F}$. We show explicitly that the
  time at which SRS begins to dominate is given almost
  exclusively by the optical depth but that the Fresnel number
  $\mathcal{F}$ is important for determining the total amount of light
  radiated from the ensemble.  Our theory is based on a generalization
  of the one dimensional theory presented in Ref.
  \cite{raymer-mostow}, but includes several effects omitted in the
  three dimensional generalization in Ref.  \cite{raymer-mostow2}. 
  The main difference is that we go beyond the extreme paraxial
  approximation used there, a generalization only briefly discussed in an
  appendix of Ref. \cite{raymer-mostow2}. The theory that we develop can
  therefore explain both spontaneous emission as well as SRS.  

  The analysis begins with the basic set of equations describing the
  interaction of light with atoms. The atoms are treated as non-moving
  point particles and the radiation fields are described by the
  displaced electric field, suited for a macroscopic description of the
  system.  See e.g. Ref. \cite{martin-anders} for a discussion of this
  choice.  We will in Sec. \ref{sec:equations-motion} derive
  effective equations of motion for both the radiation field and the
  atoms.  These equations are directly comparable to the equations
  used in Ref. \cite{raymer-mostow}. Having established the equations
  of motion we will in Sec.  \ref{sec:going-from-discrete} change from
  the point particle picture to a continuous description. This again
  follows methods described in e.g. Ref. \cite{martin-anders}. In Sec.
  \ref{sec:diagonalization} we make a formal diagonalization of the
  matrix describing the interaction between atoms mediated by the
  light.  This diagonalization means that we have to find a basis that
  will simplify the interaction. In Sec.  \ref{sec:real-space-repr} we
  will look at the radiated field and see how this is evolving as the
  atoms are interacting.  Finally in Sec.  \ref{sec:intens-corr-funct}
  we look at the intensity of the radiated field and present the final
  results.  We shall in addition to the analytical results make a
  comparison with numerical calculations for the SRS starting with the
  point particle equations of motion derived in Sec.
  \ref{sec:equations-motion}. In Sec.  \ref{sec:conclusion} we
  conclude the work.

  \section{Equations of motion}\label{sec:equations-motion}
  In the electric dipole approximation the Hamiltonian describing a
  collection of atoms is given by
 \begin{align}
    \mathcal H =& \int \{ \mathcal H_{F} + \mathcal H_{I}\} d^3r +
    \mathcal H_{A},\\
    \mathcal H_{F} =& \frac{\mathbf D^2}{2\epsilon_0} +
    \frac{\mathbf B^2}{2\mu_0} \\
        \mathcal{H}_{I} =& - 
    \frac{1}{\epsilon_0} \mathbf{D}(\mathbf{r} ,t) \cdot
    \mathbf{P}(\mathbf r ,t ) \\
    \mathcal H_{A} =& \sum_j^{Atom} \sum_{n} E_n^j \sigma_{nn}^j,
  \end{align}
  where $\mathbf D$ is the displaced electric field, $\mathbf B$ is
  the magnetic field and $\mathbf P$ is the atomic polarization.  The
  operator $\sigma_{nn}^j=|n\>\<n|$ is a projection operator for the
  $j$'th atom, and $E^j_n$ is the energy corresponding to the
  state $|n\>$.  We choose to use the displaced electric field and not
  the electric field for reasons discussed e.g. in Ref.
  \cite{martin-anders}. This choice, however, does not influence the
  result of the analysis. Here we have ignored any direct interaction
  between the atoms, e.g. atomic collisions. As we shall often make
  reference to Ref. \cite{raymer-mostow}, we shall try and match the
  constants and the dynamics of our system to the system presented
  there. The Hamiltonian is also chosen such that results derived in
  Ref. \cite{martin-anders} can be directly incorporated. In
  the following section we will focus on the dynamics of the atoms.

  \subsection{Atomic dynamics}
\begin{figure}
    \centering
    \includegraphics[width=0.3\textwidth]{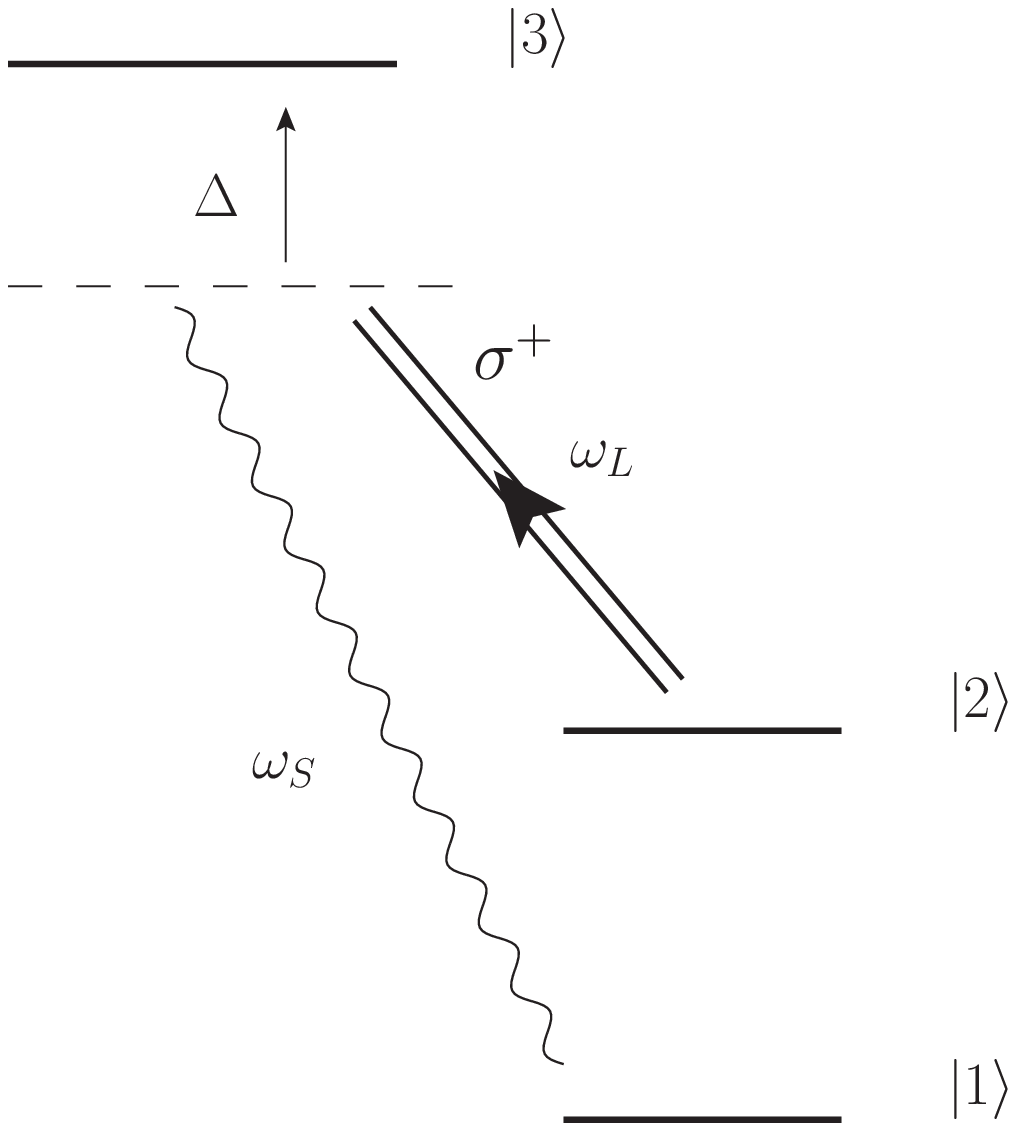}
    \caption{Atomic level structure. Two stable ground states $|1\>$
      and $|2\>$ are coupled through an exited state $|3\>$. We assume
      a strong classical laser of $\sigma^+$-polarized light drives
      the transition from $|2\>$ to $|3\>$ with detuning 
      $\Delta$. The laser thereby effectively drives a transition from
      level $|2\>$ to $|1\>$. The radiation $\stokes$ connected to the
      transition from $|3\>$ to $|1\>$ describes the Stokes field,
      that is analyzed here. }
    \label{fig:atom-level-struct}
  \end{figure}

  The macroscopic description of the atomic ensemble is given by the
  polarization, $\mathbf P(\mathbf r,t)$ which again is the sum of the
  individual dipole moment of the atoms. 
  \begin{align}
  \mathbf P(\mathbf r ,t) = \sum_j^{\text{\tiny{Atoms}}} \sum_{nm} 
  \delta(\mathbf r- \mathbf r_j) \mathbf d_{nm} \sigma^j_{nm}(t) ,
  \end{align}
  where the time dependent operator $\sigma_{nm}^j(t)$ is the operator
  $|n\>\<m|$ taking the $j$'th atom from state $|m\>$ to state $|n\>$,
  and the dipole moment is $\mathbf d_{nm}=e\< n|\mathbf r |m\>$. In
  addition we assume the atoms to be identical with a level structure
  shown in Fig.  \ref{fig:atom-level-struct}. We assume the two levels
  $|1\>$ and $|2\>$ to be stable ground states.
  For the chosen atomic system we assume that the transition from level
  $|1\>$ or $|2\>$ to $|3\>$ increases the atomic angular momentum by
  one unit of $\hbar$, and that there are no other states that the
  level $|3\>$ can decay to.
  This means that the only non-vanishing vector components of the
  dipole moments are $\mathbf e_+=(\mathbf e_x +i \mathbf e_y)/\sqrt{2}$
  for positively oscillating terms and $\mathbf e_+^*$ for negatively
  oscillating terms.

  We employ the Rotating Wave Approximation (RWA) and assume that
  $\Delta$ is sufficiently large so that we may adiabatically
  eliminate the exited level $|3\>$. In this process we split the
  radiation field $\mathbf D$ into its positively and negatively
  oscillating parts, and extract the strong classical field
  $\mathbf{\mathcal D}_{cl}$ oscillating with a frequency $\laser$
  from the weak quantum mechanical stokes field $\hat{\mathbf D}$
  oscillating with frequency $\stokes$. We will assume that the strong
  classical field is constant over the region of the atoms and can be
  written as a plane wave with a constant amplitude $\mathcal
  D_{cl}^{(+)}=|\mathcal D_{cl}|e^{-i\laser t+i\klaser z}\mathbf e_+$.
  The presence of the strong classical field $\mathbf{\mathcal
    D}_{cl}$ induce a Stark shift of the atomic levels. The effective
  Stokes frequency $\stokes$ is therefore given by
  \begin{align}
    \stokes=\laser + \omega_{21} - \frac{ |d_{31}|^2|\mathcal D_{cl}|^2 }{\hbar^2
      \epsilon_0^2 \Delta}.
  \end{align}
  We define slowly oscillating operators both for the atomic
  operator $\sigma_{21}$ and for the stokes field $\hat{\mathbf D}$
  \begin{align}
    \tilde{\sigma}_{12}(t)=&\sigma_{12}e^{i(\stokes-\laser)t-i\klaser z} \\
    \tilde{\mathbf D}^{(+)}=&\hat{\mathbf D}^{(+)}e^{i\stokes t}.
  \end{align}
  For large detuning and weak fields we can adiabatically eliminate
  the exited state, and obtain an effective ground state equation of
  motion.
  \begin{align}\label{eq:fracddts-=-frac}
    \frac{d}{dt}\tilde{\sigma}_{12}^j(t) = \frac{-i a }{\epsilon_0 \hbar
    } (\sigma^j_{22}-\sigma^j_{11})|\mathcal
    D_{cl}| \tilde D^+_-(\mathbf r_j,t), 
  \end{align}
  where the constant $a$ is given by
  \begin{align}
    a=\frac{d_{32}d^*_{31}}{\hbar \epsilon_0 \Delta}.
  \end{align}
  The positively oscillating part of the polarization is in this
  approximation
  \begin{align}\label{eq:mathbf-p+mathbf-r}
    \tilde{\mathbf P}^{(+)}(\mathbf r,t) = \sum_j a|\mathcal D_{cl}| \mathbf
    e_{+} \tilde{\sigma}_{12}^j(t) \delta (\mathbf r -\mathbf r_j).
  \end{align}
  The negatively oscillating part $\tilde{\mathbf P}^{(-)}(r,t)$ is found by
  Hermitian conjugation.

  \subsection{Field equation}
 The equation of motion for the electric field $\mathbf D(\mathbf r,t)$
 is e.g. given in Ref. \cite{martin-anders} by 
 \begin{align}\label{eq:d+mathbf-r-t}
   \mathbf D^{(+)}(&\mathbf r ,t) = \mathbf{D}_0^+(\mathbf r,t) + 
   \notag \\ &\sum_j \int dt'\: \bar{\bar{P}}^{(+)}(\mathbf r,t|\mathbf r_j,t')
   \cdot \mathbf e_+ a |\mathcal D_{cl}| \tilde{\sigma}_{12}^{j'}(t'),
 \end{align}
 where $\mathbf D_0$ is the unperturbed field containing the vacuum
 Stokes field and the classical laser-field, and $\bar{\bar P}^{(+)}$
 is the propagator. The coupling between level $|2\>$ and $|3\>$ in
 principle give rise to an index of refraction.  As shown in Ref.
 \cite{martin-anders}, such an index of refraction should 
 be incorporated into the propagator $\bar{\bar P}^{(\pm)}$.  In the
 limit of large detuning $\Delta$ (but fixed $a|\mathcal D_{cl}|$), we
 can however neglect this, and will do so in the following. The
 propagator in the slowly varying approximation is in Fourier
 representation given by
\begin{align}
  \label{eq:fouriertrans_propagator2}
  \bar{\bar P}^{(+)}(\mathbf r,\mathbf r') = \kstokes^3 \int d^3k
  \sum_{\boldsymbol{\varepsilon}\perp \mathbf k} \frac{k^2
    e^{i\mathbf k \cdot (r-r')}}{(2\pi)^3(k^2 - 1 )}
  \boldsymbol{\varepsilon}\boldsymbol{\varepsilon}^*,
\end{align} 
where the $\mathbf k$-integral is understood to include only the
contribution corresponding to the retarded Green function. Here and in
the remainder of this work we will measure the spatial coordinates in
units of $\kstokes$, which gives the factor of $\kstokes^3$ and a pole
at $1$ in Eq.  (\ref{eq:fouriertrans_propagator2}).

Inserting Eq. (\ref{eq:d+mathbf-r-t}) into
Eq. (\ref{eq:fracddts-=-frac}) gives us an effective equation of
motion for the atomic operators, 
\begin{align}\label{eq:fracddt-sigma_12j-t}
  \frac{d}{dt} \tilde{\sigma}_{12}^j (t) = - \frac{\Gamma}{2} \tilde{\sigma}_{12}^j (t)
  + \sum_{j'\neq j} M_{jj'} \tilde{\sigma}_{12}^{j'}(t) + \hat F_j(t) ,
\end{align}
where
\begin{align}
  \Gamma =& \frac{a^2 \klaser ^3 |\mathcal D_{cl} |^2}{3 \pi
    \epsilon_0 \hbar }, \\
  \hat F_j(t) =& \frac{-i a}{\epsilon_0 \hbar }\mathbf
  D_0^{(+)}(\mathbf r_j,t) \cdot \mathbf e_+^*
  \mathcal D_{cl}^{(-)}(t), \\
  M_{jj'} =& \frac{-3\pi i \Gamma}{\kstokes ^3} \mathbf e_+^* \cdot
  \bar{\bar P}^{(+)}(\mathbf r_j,\mathbf r_j') \cdot \mathbf e_+
  .\label{eq:m_jj-=-frac}
\end{align}
We have in addition made the approximation $\sigma_{22}-\sigma_{11}
\approx 1$, where we assume that initially all atoms are in state
$|2\>$ and that the experiment takes place on a timescale such that we may
neglect depletion of this level. To derive the decay $\Gamma$ we used the identity
\begin{align}\label{eq:c2-mathbf-e}
  \mathbf e_+^* \cdot \bar{\bar P}^{(+)}(\mathbf r_j \mathbf r_j )
  \cdot \mathbf e_+ = \frac{i \kstokes^3}{6\pi},
\end{align}
which is discussed in e.g. Ref. \cite{martin-anders} as the infinitely
short propagator, and the relation
$(\sigma_{22}-\sigma_{11})\sigma_{12}=-\sigma_{12}$.  The effective
equation of motion for the atoms, (\ref{eq:fracddt-sigma_12j-t})
is the starting point for many studies of SRS
\cite{raymer-mostow,akkermans}, but also for studies of the coupling
between atomic spin-excitations and collective emission of light,
\cite{cirac,hjortshoj,kurizki}.  In our analysis we neglect the effect
of the source term $\hat F_j$ in Eq.  (\ref{eq:fracddt-sigma_12j-t}),
as we are eventually only interested in measuring the photon flux
$\<D^{(-)}D^{(+)}\>$. It can be found from Eqs.
(\ref{eq:d+mathbf-r-t}) and (\ref{eq:fracddt-sigma_12j-t}) that the
effect of the source term $\hat F_j$ leads to a contribution
$\<D_0^{(-)}D_0^{(+)}\>$ to the measurement. This contribution vanish
as we assume that the Stokes field is in the vacuum state. We also
assume that there is no classical noise in the laser field $\mathcal
D_{cl}$.

We shall be interested in defining creation and annihilation
operators for the atoms. This leads in general to nonlinear equations,
but under the low excitation approximation, that is
$\sigma_{22}^j-\sigma_{11}^j \approx 1$, we employ the
Holstein-Primakoff approximation and simply use
\begin{align}
  \hat b^{\dagger}_j = \sigma_{12},\qquad \hat b_j = \sigma_{21},
\end{align}
so that 
\begin{align}
  \big[ \hat b_j , \hat b^{\dagger}_{j'} \big] = \delta_{jj'}.  
\end{align}
The effective equation of motion for the atoms is then given by 
\begin{align}\label{eq:fracddt-hat-bdagg-1}
  \frac{d}{dt} \hat b^{\dagger}_j (t) = - \frac{\Gamma}{2} \hat b_j^{\dagger} (t)
  + \sum_{j'\neq j} M_{jj'} \hat b^{\dagger}_{j'}(t),
\end{align}
and for the field Eq. (\ref{eq:d+mathbf-r-t}) gives
\begin{align}\label{eq:mathbf-d+mathbf-r}
   \mathbf D^{(+)}(&\mathbf r ,t) = \mathbf{D}_0^+(\mathbf r,t) +
   \notag \\ &
   \sum_j \int dt'\: \bar{\bar{P}}^{(+)}(\mathbf r,t|\mathbf r_j,t')
   \cdot \mathbf e_+ a |\mathcal D_{cl}| \hat b_j^{\dagger}(t).
\end{align}

\section{Going from discrete to continuous system}
\label{sec:going-from-discrete}

We will be interested in treating Eq.  (\ref{eq:fracddt-hat-bdagg-1})
as a continuous equation. For an atomic gas we do not know the
individual positions of the atoms, thus an expectation value of a
physical operator has to be accompanied by a spatial average of the
individual atomic positions. We therefore define the density
distribution $\check{\rho}(\mathbf r)$,
\begin{align}
  \check{\rho}(\mathbf r)=\sum_j \delta(\mathbf r-\mathbf r_j).
\end{align}
We assume that after a spatial average of the position of the atoms in the ensemble
the density distribution $\check{\rho}(\mathbf r)$ can be described by
a Gaussian function
\begin{align}\label{eq:rhomathbf-r=rho_0-e-1}
  \< \check{\rho}(\mathbf r)\>_{sa.}\equiv \rho(\mathbf r)=\rho_0
  e^{-\frac{r^2}{2\sigma_{\perp}^2}-\frac{z^2}{2\sigma_{||}^2}}.
\end{align}
We will also assume that $1\ll \sigma_{\perp} \ll \sigma_{||}$ and 
$\sigma_{\perp}^2>\sigma_{||}$ where spatial coordinates are measured
in units of $\kstokes$.  We then define the normalized
continuous operator
\begin{align}
  \hat b(\mathbf r) = \frac{1}{\sqrt{\rho(\mathbf r)}} \sum_j \delta
  (\mathbf r -\mathbf r_j) \hat b_j.
\end{align}
After taking spatial average of the position of the atoms, this
definition leads to the standard commutation relations for such
continuous operators,
\begin{align}
   \big[ \hat b(\mathbf r) , \hat b^{\dagger}(\mathbf r) \big] = \delta(\mathbf
   r-\mathbf r').
\end{align}
From this definition of the continuous operators Eq.
(\ref{eq:fracddt-hat-bdagg-1}) can be rewritten
\begin{align}\label{eq:fracddtbd-r-t=int}
  \frac{d}{dt}b^{\dagger}(\mathbf r,t)&=\int d^3r\: \sum_j
  \frac{\delta(\mathbf r-\mathbf r_j)}{\sqrt{{\rho} (\mathbf r)}}
  M(\mathbf r, \mathbf r') \sqrt{{\rho} (\mathbf r') }
  b^{\dagger}(\mathbf r',t) \notag\\ & =\int d^3r\: \sqrt{{\rho}
    (\mathbf r)} M(\mathbf r, \mathbf r') \sqrt{{\rho} (\mathbf r') }
  b^{\dagger}(\mathbf r',t) \notag \\ & \hspace{-1cm}+\int
  d^3r\:\sum_j \frac{\delta(\mathbf r-\mathbf r_j) -\rho(\mathbf
    r)}{\sqrt{{\rho}(\mathbf r)}} M(\mathbf r, \mathbf r')
  \sqrt{{\rho} (\mathbf r') } b^{\dagger}(\mathbf r',t).
\end{align}
The lowest order spatial average is found simply by making a spatial
average of Eq. (\ref{eq:fracddtbd-r-t=int}).  In Ref.
\cite{martin-anders} we considered higher order corrections coming
from such a spatial average, and showed how fluctuations in position
give rise to spontaneous emission and dipole-dipole interaction
effects. Here we shall ignore these effects.  To lowest order in the
spatial average, the first term in Eq.  (\ref{eq:fracddtbd-r-t=int})
describes the mean effect of the atoms interaction with each other,
that is when averaged with respect to their individual positions. The
second term will after spatial averaging only give a contribution for
atoms interacting with themselves via the infinitely short propagator
\cite{martin-anders}, thus the term effects in the decay described by
$\Gamma$, which is independent of the interactions between atoms. To
get the sign of the decay, one would have to remember that the
approximation $\sigma_{22}-\sigma_{11}\approx 1$ is not justified for
this particular type of term, and including this correction as in Eq.
(\ref{eq:fracddt-sigma_12j-t}), gives the negative sign.  The
continuous version of Eq. (\ref{eq:fracddt-hat-bdagg-1}) is then
\begin{align}\label{eq:fracddtbd-r-t=int-2}
     \frac{d}{dt}b^{\dagger}(\mathbf r,t)=\int d^3r\: \sqrt{\rho
       (\mathbf r)} M(\mathbf r, \mathbf r') &\sqrt{\rho (\mathbf r') }
     b^{\dagger}(\mathbf r',t) \notag \\ & -\frac{\Gamma}{2}b^{\dagger}(\mathbf r,t).
\end{align}
It is convenient to remove the last term  of
Eq. (\ref{eq:fracddtbd-r-t=int-2}) by defining new atomic operators
with respect to the decay $\Gamma$, [ $\hat
b^{\dagger}(\mathbf r,t) \rightarrow \hat b^{\dagger}(\mathbf
r,t)e^{-\frac{\Gamma}{2} t} $ ], ignoring the source term $\hat F$ and the
point-particle corrections, the effective differential equation
describing the excitation of the atoms is after spatial average
given by
   \begin{align}\label{eq:fracddtbd-r-t=int-1}
     \frac{d}{dt}b^{\dagger}(\mathbf r,t)=\int d^3r\: \sqrt{\rho
       (\mathbf r)} M(\mathbf r, \mathbf r') \sqrt{\rho (\mathbf r') }
     b^{\dagger}(\mathbf r',t).
   \end{align}
Similarly the field equation (\ref{eq:d+mathbf-r-t}) can be described in
terms of the continuous operators, and one find
\begin{align}\label{eq:mathbf-d+mathbf-r-1}
  \mathbf D^{(+)}(&\mathbf r ,t) = \mathbf{D}_0^+(\mathbf r,t) +
  \notag \\ & a |\mathcal D_{cl}| \int d^3 r'
  \bar{\bar{P}}^{(+)}(\mathbf r,\mathbf r')\sqrt{\rho(\mathbf r')}
  \cdot \mathbf e_+ \hat b^{\dagger}(\mathbf r',t).
\end{align}
In the following we will find approximate solutions to the above
equations.

\section{Diagonalizing the interaction
  matrix}\label{sec:diagonalization}

\begin{figure}
  \centering
  \includegraphics[width=0.48\textwidth]{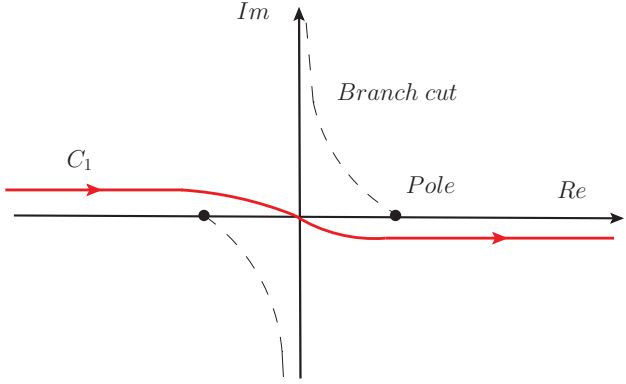}
  \caption{Sketch of the integration contour $C_1$, in the integral
    representation (\ref{eq:frac-r-mathbf}) of the Green function. }
  \label{fig:plot-curve-c_1}
\end{figure}

The system is assumed to be cylindrically symmetric, with a
density described by Eq.
(\ref{eq:rhomathbf-r=rho_0-e-1}).  We shall therefore use a
cylindrically symmetric set of basis functions for our
diagonalization: a combination of plane waves and Bessel functions. We
denote the basis by $\{ f_{kmn}\}$, where
\begin{align}
  f_{kmn}(r,z,\phi)=\frac{\sqrt{2}}{2\pi a_c
    J_{m+1}(X_{mn})}e^{ikz+im\phi}J_{m}(X_{mn}\frac{r}{a_c}).
\end{align}
$J_m$ is the Bessel function of first kind of order $m$, and $X_{mn}$
is the $n$'th zero of the $m$'th order Bessel function of first kind.
The parameter $a_c$ is a cut-off in the radial direction, meaning that
our basis is complete on the interval $r\in [ 0, a_c] $.  The inner
product defined for this basis is therefore given by,
\begin{align}
  \< \theta | \psi \> = \int_0^{2\pi} d\phi \int_{-\infty}^{\infty}  dz
  \int_0^{a_c} rdr \; \theta^*(r,z,\phi) \psi(r,z,\phi). 
\end{align}
For a discussion of this basis see e.g. Ref. \cite{jackson}. To solve
Eq. (\ref{eq:fracddtbd-r-t=int-1}) we will diagonalize the matrix
given by
\begin{align}
  M(\mathbf r,\mathbf r') = \frac{-3\pi i \Gamma}{\kstokes^3} \mathbf
  e_+^* \cdot \sqrt{\rho(\mathbf r)}\bar{\bar P}^{(+)}(\mathbf
  r,\mathbf r')\sqrt{\rho(\mathbf r')} \cdot \mathbf e_+.
\end{align}
The propagator $\bar{\bar P}^{(+)}$ is found in a real space
representation in e.g. Ref. \cite{morice}. One may from the real
space representation of the propagator show that
\begin{align}\label{eq:mathbf-e_-cdot}
  \mathbf e_+^* \cdot
\bar{\bar{P}}^{(+)}(\mathbf r,\mathbf r') \cdot \mathbf e_+ =
\frac{-\kstokes^3}{8\pi}\big( \nabla^2 +\partial_z^2 \big)
\frac{e^{i|\mathbf r - \mathbf r'|}}{|\mathbf r - \mathbf r'|}.
\end{align}
The polarization effects are here included in the differential operator
$\nabla^2 +\partial_z^2$.  In addition we use 
that the Green's function can be written as  \cite{klimow}
\begin{align}\label{eq:frac-r-mathbf}
  \frac{e^{i|\mathbf r - \mathbf r'|}}{|\mathbf r - \mathbf
    r'|}=\frac{i}{2}\sum_m\int_{C_1}& dh e^{im(\phi-\phi')+ih(z-z')}
 \notag \\ &J_m(\sqrt{1-h^2}r_{<}) H^{(1)}_m(\sqrt{1-h^2}r_{>}),
\end{align}
where $r_{<}\;(r_{>})$ is the minor (larger) of $r$ and $r'$. $C_1$
is describing a curve essentially going from $-\infty$ to $\infty$
along the real axis but shifted to avoid the branch cut and pick out
the retarded Green's function, as shown in Fig.
\ref{fig:plot-curve-c_1}.
By introducing an integral, the non-trivial product of Bessel
functions in Eq. (\ref{eq:frac-r-mathbf}), can be symmetrized \cite{integrals}:
\begin{align}
  J_m(\sqrt{1-h^2}r_{<}) H^{(1)}_m(&\sqrt{1-h^2}r_{>}) \notag \\ &=  
  \frac{2}{i\pi}\int xdx \frac{J_m(xr)J_m(xr')}{x^2+h^2-1}. 
\end{align}
 The propagator is then given by 
 \begin{align}\label{eq:mathbf-e_-cdot-1}
   \mathbf e_+^* \cdot
\bar{\bar{P}}^{(+)}(\mathbf r,\mathbf r') \cdot \mathbf e_+ =&
\frac{\kstokes^3}{8\pi^2}\sum_m\int_{C_1}dh\int xdx
\frac{1+h^2}{x^2+h^2-1} \notag \\ &
e^{im(\phi-\phi')+ih(z-z')}J_m(xr)J_m(xr'). 
 \end{align}
In the basis $\{ f_{kmn} \}$ the differential equation
(\ref{eq:fracddtbd-r-t=int-1}) can be written
\begin{align}
  \frac{d}{dt}\hat b^{\dagger}_{kmn}(t) =\sum_{k'm'n'}
  M^{kmn}_{k'm'n'} \hat b^{\dagger}_{k'm'n'} 
\end{align}
where 
\begin{align}\label{eq:mkmn_kmn=-f_kmnm-r}
  M^{kmn}_{k'm'n'}=&\< f_{kmn}(\mathbf r) | M(\mathbf r,\mathbf r') |
  f_{k'm'n'}(\mathbf r') \>, \intertext{and}
  b^{\dagger}_{kmn}(t) =&\<f_{kmn}(\mathbf r) | b^{\dagger}(\mathbf r,t)\>.
\end{align}
When calculating the matrix Eq. (\ref{eq:mkmn_kmn=-f_kmnm-r}), we have
to make integrals over $r,z$ and $\phi$. We can at this point simplify
the radial integrals by extending the upper integral limit to
infinity. This is correct since the cut-off $a_c$ can be chosen
arbitrarily and as we in the end will set it to infinity. Due to
finite width $\sigma_{\perp}$ of the density function, this limit
accurately describe the matrix elements for $a_c \gg \sigma_{\perp}$. 
After making the spatial integrations the matrix $M$ reduces to 
\begin{widetext}
 \begin{align}\label{eq:m_kmnkmn-=-delta_mm}
    M_{kmn}^{k'm'n'} =& \delta_{mm'} \frac{\lambda_0}{i} \int_{C_1} dh
    \int xdx \; \eta(k-h) \eta(k'-h) \frac{1+h^2}{x^2+h^2-1}
    \frac{8\sigma_{\perp}^4 e^{-\sigma_{\perp}^2(\gamma_n^2
       +\gamma_{n'}^2)}}{a_c^2 J_{m+1}(X_{mn}) J_{m+1}(X_{mn'})}\times 
    \notag \\ & \hspace{4cm} e^{-2\sigma_{\perp}^2x^2} I_m(2\sigma_{\perp}^2\gamma_{n}x)
    I_m(2\sigma_{\perp}^2\gamma_{n'}x)
 \end{align}
\end{widetext}
where 
\begin{align}
    \eta(k)=& \frac{\sigma_{||}}{\sqrt{\pi}}e^{- \sigma_{||}^2k^2},
\end{align}
and where we have introduced the constant $\lambda_0=\frac{3 \pi
  \rho_0 \Gamma}{2}$. To shorten notation we also introduce
$\gamma_n=\frac{X_{mn}}{a_c}$, where we understand that $\gamma_n$
depend on the azimuthal quantum number $m$. For integrals involving Gaussian functions and
Bessel functions we refer to Ref. \cite{integrals}.  We notice that
both integrals over $x$ and $h$ are bounded by Gaussian functions, and
since we assume $\sigma_{\perp}\gg 1$ we may make a series expansion
in $x$ and $h$ of the function $1/(x^2 +h^2 -1)$. We will be
interested in a series expansion of the integrals over $x$ and $h$
only to the lowest order. Since we assume that $\sigma_{||} \gg
\sigma_{\perp}$, i.e.  cigar-shape, our lowest order calculation will
terminate after first order in $1/\sigma_{\perp}^2$. The integral over
$h$ can to this order be approximated by treating the function
$\eta(k-h)$ as a delta function, thus we shall here and in the
remainder of the article treat the function $\eta(k-h')$ as a delta
function. We show in Appendix \ref{sec:deriving-first-order} that the
integral over $x$ to lowest order in the variable $1/\sigma_{\perp}^2$
gives
\begin{widetext}
  \begin{align}\label{eq:m_kmnkmn-=-delt}
    M_{kmn}^{k'm'n'} =& \delta_{mm'}\eta (k-k') \frac{\lambda_0}{ i} 
     \left\{  \Lambda_{nn'}^m  \frac{1+ k^2} {k^2 -1} - \frac{
         {\Lambda^1}_{nn'}^m }{\sqrt 8 \sigma_{\perp}^2}
       \frac{1+k^2}{(k^2 -1)^2} 
  \right\}  +  O\left[\sigma_{||}^{-2},\sigma_{\perp}^{-4}\right],
  \end{align}
\end{widetext}   
where
 \begin{align}\label{eq:lambd-=-frac2s-1}
   \Lambda_{nn'}^m =& \frac{2\sigma_{\perp}^2
     e^{-\frac{\sigma_{\perp}^2}{2}(\gamma_n^2+ \gamma_{n'}^2)}
     I_m\big( \sigma_{\perp}^2\gamma_n\gamma_{n'}\big)}{a_c^2
     J_{m+1}(X_{mn}) J_{m+1}(X_{mn'})}
  \end{align}
and 
\begin{align}\label{eq:lambd-=-frac4s}
  {\Lambda^1}_{nn'}^m =& \frac{4\sigma_{\perp}^2
     e^{-\sigma_{\perp}^2(\gamma_n^2+ \gamma_{n'}^2)}
     I_m\big(2 \sigma_{\perp}^2\gamma_n\gamma_{n'}\big)}{a_c^2
     J_{m+1}(X_{mn}) J_{m+1}(X_{mn'})}.
\end{align}
The matrices $\Lambda^m_{nn'}$ and ${\Lambda^1}^m_{nn'}$ are
normalized such that for $\sigma_{\perp}\rightarrow \infty$ they
reduce to a delta-function $\delta(n-n')$.

In the following we take a closer look at the matrix $\Lambda_{nn'}^m$
defined in Eq. (\ref{eq:lambd-=-frac2s-1}). For simplicity we will not
consider the correction ${\Lambda^1}^m_{nn'}$, however the conclusions
drawn in the following holds for the correction as well.  The
differential equation for our system with respect to the quantum
number $n,n'$ has got the form
\begin{align}\label{eq:fracddt-a_nt-=}
  \frac{d}{dt} b_n(t) = \sum_{n'} i \Omega \Lambda_{nn'}^m b_{n'}(t),
\end{align}
where $\omega$ is some real number.
We wish to take the limit $a_c\rightarrow \infty $. To clarify what
this means let us write the matrix $\Lambda$ in the following way:
\begin{align}
  \Lambda_{nn'}^m =\;& \Delta k_{mn'}  \Xi^m_{nn'} \pi^2
  \sigma_{\perp}^2 e^{-\frac{\pi^2 \sigma_{\perp}^2}{2} ( k_{mn}^2 +
    k_{mn'}^2) } \notag \\ &I_m(\pi^2 \sigma_{\perp}^2 k_{mn}k_{mn'})
  \sqrt{k_{mn}k_{mn'}}  
\end{align}
where
\begin{subequations}
  \begin{align}
    \Xi^m_{nn'}=\;& \frac{2}{\pi} \frac{1}{\sqrt{X_{mn} X_{mn'}}
      J_{m+1}(X_{mn}) J_{m+1}(X_{mn'})} \\ \approx \;& (-1)^{n+n'}
    \quad \text{for}\quad X_{mn },X_{mn'}\rightarrow \infty \notag
    \\
    k_{mn}=\;&\frac{X_{mn}}{\pi a_c}  \\
    \Delta k_{mn'}=\;&\frac{1}{a_c}
  \end{align}
\end{subequations}
We thus see that when letting $a_c\rightarrow \infty$, a transverse
momentum naturally arises $k_{m\perp}=\lim_{a_c \rightarrow \infty}
k_{mn}$, and the discrete matrix equation,
Eq. (\ref{eq:fracddt-a_nt-=}) becomes an integral equation over the
transverse momentum $k_{m\perp}$, using $\sum_{n'} \Delta k_{mn'}
\rightarrow \int dk_{m\perp}$ 
\begin{align}\label{eq:fracddtgrsdsg}
  \frac{d}{dt} b(k_{m\perp} t) = \int dk_{m\perp}' \Omega
  \Lambda^m(k_{m\perp},k_{m\perp}')  b(k_{m\perp}',t).
\end{align}
It is evident that when using the
limiting properties of the Bessel function $I_m(x)$ the integral
kernel  $\Lambda^m(k_{m\perp},k_{m\perp}')$ becomes a delta function
for $\sigma_{\perp}\rightarrow \infty$. 
\begin{align}
  \pi^2& \sigma_{\perp}^2 e^{-\frac{\pi^2 \sigma_{\perp}^2}{2} (
    k_{m\perp}^2 + {k'}_{m\perp}^2) } I_m({\scriptsize{\pi^2
      \sigma_{\perp}^2 k_{m\perp}k'_{m\perp}}})
  \sqrt{k_{m\perp}k'_{m\perp}} \notag \\ & \approx\frac{1}{\sqrt{\pi}}
  \sqrt{\frac{\pi^2 \sigma_{\perp}^2}{2}} e^{-\frac{\pi^2
      \sigma_{\perp}^2}{2} ( k_{m\perp} - k'_{m\perp})^2 } \rightarrow
  \delta(k_{m\perp} - k'_{m\perp})
\end{align}
We thus realize that the effective one-dimensional result obtained by
Raymer and Mostowski \cite{raymer-mostow} is exact for every
transverse mode in an infinitely wide atomic ensemble. In this limit
however there is no limitations on the transverse momentum, which
results in an infinite intensity. To obtain finite results we thus
need to consider the full solution to the three dimensional problem.

Now we again include the correction ${\Lambda^1}^m_{nn'}$ in the
analysis. Both matrices $\Lambda^m_{nn'}$ and ${\Lambda^1}^m_{nn'}$
are real and symmetric and can thus be diagonalized. In Appendix
\ref{sec:comm-relat-lambd} we show that the two matrices commute. We
can therefore choose a common set of eigenfunctions, $\Big\{
F_{kmn}(r)\Big\}$ for both matrices.  We define the unitary matrix
$\bar{\bar U}$ that transform our initial basis $\{ f_{kmp} \}$ to the
basis given by the eigenfunctions $\Big\{ F_{kmn}(r)\Big\}$,
\begin{align}\label{eq:f_kmnmathbf-r-=}
  F_{kmn}(\mathbf r) = \sum_{p} U_{np}f_{kmp}(\mathbf r)
\end{align}
Finally we will define a corresponding set of eigenvalues, 
\begin{align}\label{eq:lambd-u_np-lambd}
  \Lambda_{pp'}^m=\sum_n {U^{\dagger}}_{pn} \lambda_{mn} U_{np'}  
\end{align}
and 
\begin{align}\label{eq:lambd-u_np-lambd-1}
  {\Lambda^1}_{pp'}^m=\sum_n {U^{\dagger}}_{pn} \lambda^1_{mn} U_{np'}  
\end{align}
 It is convenient in the following to change to this basis, where
$\Lambda_{nn'}^m$ and ${\Lambda^1}_{nn'}^m$ are diagonal. We therefore write
Eq.~(\ref{eq:m_kmnkmn-=-delt}) as
\begin{align}\label{eq:sum_pp-notag}
  &\sum_{pp'}{U^{\dagger}}_{pn}M^{k'm'p'}_{kmp}U_{n'p'} = \notag \\  
   &\frac{\lambda_0}{i} \left\{ \lambda_{mn} \frac{1+ k^2} {k^2 -1} -
    \frac{{\lambda^1}_{mn}}{\sqrt 8\sigma_{\perp}^2} \frac{1+k^2}{(k^2 -1)^2} \right\}
  \delta_{mm'}\delta_{nn'} \eta (k-k') \notag \\ & \hspace{3cm}+
  O\left[\sigma_{||}^{-2},\sigma_{\perp}^{-4}\right].
\end{align}

\section{Real space representation of the electric
  field}\label{sec:real-space-repr}

In the following section we will, based on the eigenvalue analysis of
the atomic operators, derive the real-space behavior of the electric
field. We shall divide the analysis into a regime of small times where
the dominating effect is spontaneous emission, and a large time
regime, where the dominating effect is the cooperatively emitted
light, the SRS beam. To keep things simple, we mainly consider the
electric field at and around the symmetry axis. In this region the
scattered radiation field is sufficiently well described by the vector
component $D^{(+)}_+$ and its Hermitian conjugate. This can be seen
from Eq. (\ref{eq:mathbf-d+mathbf-r-1}) and the real space
representation of the propagator (\ref{eq:fouriertrans_propagator2}).

Let us first determine the electric field on the symmetry axis at
the initial time, $t=0$. In this case the electric field is given by:
\begin{align}
  D^{(+)}_+(\mathbf r_{s},0) =& D^{(+)}_+(\mathbf r_{s},0)_0 + \int
  d^3r'\;\frac{a|\mathcal D_{cl}|\kstokes^3}{4\pi} \hat
  b^{\dagger}(\mathbf r',0) \times \notag \\ &
  \frac{((z-z')^2+\frac{1}{2}r^2)e^{i\sqrt{(z-z')^2+r^2}}}{(r^2+(z-z')^2)^{3/2}}
  \sqrt{\rho(\mathbf r')} ,
  \label{eq:int-d3r-}
\end{align} 
where the index $s$ refers to being at the symmetry axis.  To arrive
at the above result we used the real space representation of the
propagator $\bar{\bar{P}}$ Eq. (\ref{eq:mathbf-e_-cdot}) to leading
order in one over distance. This approximation is done out of
convenience but is not strictly necessary. When calculating the mode
expansion of the electric field in the general modes $F_{kmn}$ we
shall check that the limit $t\rightarrow 0$ exist and is given by the
expression,~(\ref{eq:int-d3r-}).

The analysis of the radiation field for $t\neq0$ starts by inserting
the identity operator,
\begin{align}
  \openone =\int d^3r' \int dk \sum_{mn} F_{kmn}(r)F^*_{kmn}(r')
\end{align}
into the field equation, (\ref{eq:mathbf-d+mathbf-r-1}). We then
get the following expansion of the electric field. 
\begin{align}
  D^{(+)}&_+(\mathbf r,t) = D^{(+)}_+(\mathbf r,t)_0 \notag \\ &+ 
  \int d^3r'\; \int dk \sum_{mn} \mathcal C_{kmn}(\mathbf r)
  e^{\lambda_{kmn}t} F^*_{kmn}(\mathbf r') \hat b^{\dagger}(\mathbf
  r',0),
\end{align}
where
\begin{align}
  \mathcal C_{kmn}(\mathbf r) =& a|\mathcal D_{cl}|\hspace{-2pt}\int
  d^3r' \mathbf e_+^* \hspace{-2pt}\cdot \hspace{-2pt}\bar{\bar
    P}^{(+)}(\mathbf r,\mathbf r') \hspace{-2pt}\cdot \hspace{-2pt}
  \mathbf e_+ \sqrt{\rho(\mathbf r')} F_{kmn}(\mathbf r'),
\end{align}
the functions $F_{kmn}$ are the basis functions given in
Eq. (\ref{eq:f_kmnmathbf-r-=}), and the eigenvalue $\lambda_{kmn}$ is
given in Eq. (\ref{eq:sum_pp-notag}).

The calculation of the modefunctions $\mathcal C_{kmn}$ is initiated
by integrating with respect to the spatial coordinate $\mathbf r'$.
The integrals involving Bessel functions are found in e.g. Ref.
\cite{integrals}, and one arrive at
\begin{widetext}
  \begin{align}\label{eq:mathc-c_kmnm-r}
    \mathcal C_{kmn}(\mathbf r) =& \frac{ a |\mathcal D_{cl}|
      \kstokes^3\sqrt{\rho_0}(1-\partial_z^2)}{4\pi} \int dy \int xdx
    \sum_p U_{np} \frac{e^{im\phi+i(k+y)z}}{x^2+(k+y)^2-1} \frac{\sqrt
      2 J_m(xr)}{a_cJ_{m+1}(X_{mp})}\times \notag \\ & \hspace{5cm}
    \frac{2\sigma_{\perp}^2\sigma_{||}}{\sqrt{\pi}}
    e^{-\sigma_{||}^2y^2-\sigma_{\perp}^2(\gamma_p^2+x^2)}
    I_m(2\sigma_{\perp}^2\gamma_px).
  \end{align}
\end{widetext}
The next step of the calculation is to include the mode summation. We
will therefore define the propagator $P^{(+)}$ given by
\begin{align}\label{eq:int-dk-sum_mn}
  P^{(+)}(\mathbf r,\mathbf r';t)=\int dk\; \sum_{mn} \mathcal
  C_{kmn}(\mathbf r)e^{\lambda_{kmn}t} F^*_{kmn}(\mathbf r').
\end{align}
We notice that the variable $y$ in Eq.  (\ref{eq:mathc-c_kmnm-r}) is
small, as it is controlled by the Gaussian function of width $1/\sigma_{||}$.
We shall therefore by a translation of the integral variable $k'=k+y$
move the perturbation $y$ to the eigenvalue $\lambda_{kmn}$, so that
we use $\lambda_{k'-y,mn}$. This choice ensure that we will get the
correct behavior of the integrals in the limit $t=0$.  By doing this
we can then in principle make the $k'$ integral by using the series
expansion of the function $e^{\lambda_{k'-y,mn}t}$, where the zeroth
order term in the expansion in $t$ is the limit given by Eq.
(\ref{eq:int-d3r-}). In order to accurately capture the exponential
growth, we however, instead follow the path used by e.g. Ref.
\cite{raymer-mostow}.

In the following we make a series expansion of the eigenvalue
$\lambda_{k-y,mn}$ given in Eq. (\ref{eq:sum_pp-notag}) with respect
to the variable $y$.
\begin{align}\label{eq:lambda_k-y-mnequiv}
  \lambda_{k-y,mn}\equiv & \frac{1}{i} \left(
    \lambda_{mn}\frac{(k-y)^2+1}{(k-y)^2-1}-\frac{\lambda^1_{mn}
    }{\sqrt 8 \sigma_{\perp}^2 } \frac{(k-y)^2+1}{((k-y)^2-1)^2} \right)  \notag \\
  \approx & \frac{1}{i} \left( \lambda_{mn}\frac{k^2+1}{k^2-1}+
    2\mu_{mn}\frac{k^2+1}{(k^2-1)^2} \right),
 \end{align}
 The series expansion can be done since the $y$-integral is bounded by a
 Gaussian function. To shorten notation we have substituted $k'\rightarrow
 k$, and introduced the coefficient $\mu_{mn}=
 \lambda_{mn}y-\frac{\lambda^1_{mn} }{\sqrt 8 \sigma_{\perp}^2}$.  

 In Eq.  (\ref{eq:int-dk-sum_mn}) the $k$-integral includes a pole
\begin{align}
 \frac{1}{k^2+x^2-1}\rightarrow
  \frac{1}{2\sqrt{1-x^2}(k-\sqrt{1-x^2})},
\end{align}
where the arrow reflects the fact that we are only interested in the
retarded Green function, which correspond to the pole
$k=\sqrt{1-x^2}$. Since we are particularly interested in this pole,
we shall in the $k$-integral in Eq. (\ref{eq:int-dk-sum_mn}), make a
translation of the eigenvalue $\lambda_{k-y,mn}\rightarrow
\lambda_{k-y+\sqrt{1-x^2},mn}$, and then a series expansion similar to
Eq. (\ref{eq:lambda_k-y-mnequiv}).  We can make the calculation with
two different situations in mind: One situation explains the
spontaneous radiation originating from a sample of atoms of some
geometrical shape. We are most interested in the other situation
describing the collective emission or the SRS occurring when the atoms
co-radiate. As a check of our formalism we shall, however, also
consider the short time-limit where there is just spontaneous
emission. We expect that as time evolves the SRS effect will become
dominant.  Therefore we demonstrate where the SRS effect is found and
described in our mathematical treatment of the problem.

Let us first show how the important steps in the calculation of
SRS is done, before going into the full details. The
integral appearing in the calculation is of the type
\begin{align}\label{eq:mathc-i_k=fr-int}
  \mathcal I_k(t)=\frac{1}{2\pi} \int dk
  \frac{e^{\lambda_{k-y,mn}t+ik\Delta z}}{k^2+x^2-1},
\end{align}
where $\Delta z=z-z'$.  For now we consider the lowest order
correction for simplicity, that is we neglect $\mu_{mn}$ in Eq.
(\ref{eq:lambda_k-y-mnequiv}). Including $\mu_{mn}$ to the eigenvalue
is a trivial generalization.  We focus on the pole in the integral at
$k=\sqrt{1-x^2}$, as this pole describes the energetically allowed
scattering processes.  By introducing the variable $s=i\Delta
z(k-\sqrt{1-x^2})$ the integral $\mathcal I_k^0$ can be written
\begin{align}\label{eq:mathcal-i_k0-=}
  \mathcal I_k^0(t) = i \frac{1}{2\pi i} \int _{-i\infty}^{i \infty} ds \frac{e^{s+i\Delta
    z\sqrt{1-x^2}+\frac{\lambda_{mn} t \Delta z}{s+i\Delta
      z(\sqrt{1-x^2} -1)}}} {2 \sqrt{1-x^2} s},
\end{align}
where the superscript $0$ indicates that this is a zeroth order
calculation in the correction to the eigenvalue due to finite size.
The SRS contribution to Eq. (\ref{eq:mathcal-i_k0-=}) comes
from the pole of the exponential. In order for this pole to contribute
to the pole describing the propagated light, that is the zero point of
the denominator, the term $\Delta z(\sqrt{1-x^2} -1)$ has to be small.
For $\Delta z(\sqrt{1-x^2} -1)<1$ we shall treat it as a perturbation.
When this no longer apply, the pole in the exponent can be neglected,
and we are thus left with the result for short times, i.e. spontaneous
emission. The latter is analyzed in the following section, and we
shall for now concern ourselves with the SRS contribution.
For reasons discussed in Sec. \ref{sec:finite-time-build} we will,
when discussing SRS, use that $\Delta z$ is large, so that
$\Delta z(\sqrt{1-x^2} -1) \approx \frac{x^2\Delta z}{2}$. Since
$\frac{x^2\Delta z}{2}<1$ we can make an expansion in this quantity
and obtain
\begin{align}
  \mathcal I_k(t)= \frac{ie^{i\Delta z}}{2} \sum_{l=0}^{\infty}&
  \sum_{q=0}^{\infty} \left( \frac{ix^2\Delta z}{2}\right)^l
  \frac{\left(-2it\mu_{mn} \Delta z^2\right)^q}{q!} \times 
   \notag \\ & \frac{1}{2\pi i} \int_{-i\infty}^{i \infty} ds
  \frac{e^{s+\frac{\lambda_{mn}t\Delta z}{s}} }{s^{1+l+2q}}.
\end{align}
Here we include the correction to the eigenvale in Eq.
(\ref{eq:lambda_k-y-mnequiv}).  The integral may be found in Ref.
\cite{besselbook} and we find
\begin{align}\label{eq:mathc-i_k=-frac}
  \mathcal I_k(t)= \frac{ie^{i\Delta z}}{2} \sum_{l=0}^{\infty}&
  \sum_{q=0}^{\infty} \left( \frac{ix^2\Delta z}{2}\right)^l
  \frac{\left(-2i\mu_{mn}\Delta z^2\right)^q}{q!} \times \notag \\ &
  \frac{I_{l+2q}(2\sqrt{\lambda_{mn}t \Delta z})}{(\sqrt{\lambda_{mn}
      t \Delta z})^{l+2q}}.
\end{align} 

\subsection{Short time limit}

In order to understand our calculation of SRS, we first analyze it for
$t=0$, as we know how the propagator for $t=0$ looks when measured on
the symmetry axis. The $t=0$ regime is also met for $\frac{x^2\Delta
  z}{2}>1$. We shall also refer to this calculation as the short time
limit. Here we find from a residue calculation Eq.
(\ref{eq:mathc-i_k=fr-int}) to give
\begin{align}\label{eq:mathc-i_kt=0=fr-eisq}
  \mathcal I_k(0)=\frac{i e^{i\sqrt{1-x^2} \Delta z}}{2\sqrt{1-x^2}}.
\end{align}
By inserting this into the propagator in Eq. (\ref{eq:int-dk-sum_mn}),
the propagator may be written
\begin{widetext}
  \begin{align}\label{eq:sum_mn-frac-mathcal}
     P^{(+)}(\mathbf r,\mathbf r';0) =\sum_{mn} \frac{ a |\mathcal D_{cl}|
      \kstokes^3\sqrt{\rho_0}(1-\partial_z^2)}{4\pi}
     \int& xdx \sum_{pp'} {U^{\dagger}}_{pn} U_{np'}
     \frac{2i\sigma_{\perp}^2 e^{im\Delta\phi+i\sqrt{1-x^2}\Delta
         z}}{\sqrt{1-x^2}}\times \notag \\ & \frac{
        J_m(xr)J_m(\gamma_{p'}r')I_m(2\sigma_{\perp}^2\gamma_px)}{a_c^2J_{m+1}(X_{mp})J_{m+1}(X_{mp'})}
            e^{-\sigma_{\perp}^2(\gamma_p^2+x^2)-\frac{{z'}^2}{4\sigma_{||}^2}},
  \end{align}
\end{widetext}
where $\Delta \phi = \phi - \phi '$.
The only dependence on the mode-index $n$ is in the product of the two
matrices $U_{np}U_{np'}$ and the sum over $n$ reduces to a delta
function $\delta_{pp'}$. We then, similar to Sec.
\ref{sec:diagonalization}, identify $\sum_p \frac{1}{a_c} \rightarrow \int
\frac{d\gamma_p}{\pi}$ for $a_c\rightarrow \infty$. The variable
$\gamma_n$ is in this sense fixed, thus letting $a_c\rightarrow \infty
$ has to be accompanied by $X_{mn}\rightarrow \infty$. Therefore we
can use the large argument approximation for the Bessel functions,
\begin{align}
  J_{m+1}(X_{mn})\approx \sqrt{\frac{2}{\pi
      X_{mn}}}\cos(X_{mn}-\frac{m\pi}{2}-\frac{\pi}{4}),\hspace{4pt} X_{mn}\gg 1.
\end{align}
Using this we can make the integrals over $\gamma_p$ and $\gamma_{p'}$. 
The result of the mode
summation (\ref{eq:int-dk-sum_mn}) is then 
\begin{widetext}
  \begin{align}\label{eq:p+mathbf-r-mathbf}
    P^{(+)}(\mathbf r,\mathbf r';0) =& \frac{ a |\mathcal D_{cl}|
      \kstokes^3 \sqrt{\rho(\mathbf r')}(1-\partial_z^2)} {8\pi}
    \sum_{m}e^{im\Delta\phi} \int xdx
    \frac{ie^{i\sqrt{1-x^2}\Delta z}} {\sqrt{1-x^2}}
    J_m(xr)J_m(xr')
    \end{align}
\end{widetext}
This is the main result of this section. To verify the validity of the
approach taken so far, we shall now show that the propagator
(\ref{eq:p+mathbf-r-mathbf}) reduces to the one found 
 on the symmetry axis, (\ref{eq:int-d3r-}).   In order to show this
we will use the summation theorem for Bessel functions, see e.g.
\cite{integrals},
\begin{align}
  \sum_m e^{im\Delta \phi} J_m(xr)J_m(xr')=J_0(xR),
\end{align}
where $R=\sqrt{r^2+{r'}^2-2rr'\cos(\Delta \phi)}$.  In this way the propagator in
Eq. (\ref{eq:p+mathbf-r-mathbf}) can be written
\begin{align}
  P^{(+)}(\mathbf r,\mathbf r',0)=&\frac{ a |\mathcal D_{cl}|
    \kstokes^3\sqrt{\rho(\mathbf r')}(1-\partial_z^2)}{8\pi} \times
  \notag \\ & \hspace{2cm}  \int xdx  
  \frac{ie^{i\sqrt{1-x^2}\Delta z}
    J_0(xR)}{\sqrt{1-x^2}}.
\end{align}
The $x$-integral is known and may be found in Ref. \cite{besselbook}, to
give
\begin{align}
    P^{(+)}(\mathbf r,\mathbf r',0)=\frac{- a |\mathcal D_{cl}|
      \kstokes^3\sqrt{\rho(\mathbf r')}(1-\partial_z^2) }{8\pi}
     \frac{e^{i\sqrt{{R}^2+\Delta z^2}}}{\sqrt{{R}^2+\Delta z^2}}.
\end{align}
Finally the $z$ differential give us the result we are looking for.
\begin{align}
    P^{(+)}(&\mathbf r,\mathbf r',0)
 =\frac{ a |\mathcal D_{cl}|
      \kstokes^3\sqrt{\rho (\mathbf r')}}{4\pi}
     \frac{e^{i\sqrt{{R}^2+\Delta z^2}}}{\sqrt{{R}^2+\Delta
         z^2}}\frac{\frac{1}{2} {R}^2+\Delta z^2}{{R}^2+\Delta
       z^2}. \label{eq:-=frac-mathcal}
\end{align}
When we then look at the symmetry axis, the variable $R$ reduce to
$r'$ and we are left with the result in Eq. (\ref{eq:int-d3r-}).
The result of this section can be written as 
\begin{align}\label{eq:d+_-mathbf-r-1}
  D^{(+)}_+(\mathbf r,0) =& D^{(+)}_+(\mathbf r,0)_0 +
  \int d^3r'\; P^{(+)}(\mathbf r,\mathbf r';0) \hat b^{\dagger}(\mathbf
  r',0).
\end{align}

\subsection{ Finite time, build up of SRS }
\label{sec:finite-time-build}
\begin{figure}
  \centering
  \includegraphics[width=.48\textwidth]{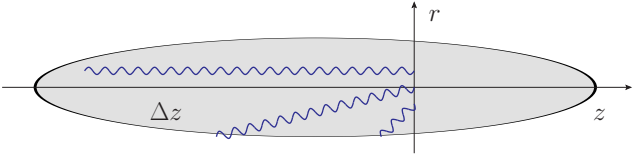}
  \caption{A sketch of the coherent build-up of radiation in an atomic
    cloud. In principle the build-up can happen along any direction,
    however for a cigar-shaped geometry the most
    significant build-up happens along the axis of the cigar.}
  \label{fig:draw-coher-build}
\end{figure}

In the following we shall analyze the effect of the eigenvalues
$\lambda_{mn}$ and $\lambda^1_{mn}$ in the expression
(\ref{eq:mathc-i_k=-frac}). When we introduced the eigenvalues in
Sec. \ref{sec:diagonalization} we only concluded they could be
found. We also know that physics connected to the eigenvalues can not depend
on the cut-off $a_c$ involved in the index $n$. In the following we show that
indeed the physics is independent of the cut-off $a_c$.
To find this result we shall in particular look at the sum $\sum_n
{U^{\dagger}}_{pn}\mu_{mn}^M\lambda^N_{mn} U_{np'}$ where the powers $N$ and $M$
are zero or some positive integer. [ The powers $N$ and $M$ are
connected to the series expansions of functions involving the eigenvalue
$\lambda_{mn}$, e.g. Eq.  (\ref{eq:mathc-i_k=-frac}). ]  $\lambda_{mn}$
and $\lambda^1_{mn}$ are the eigenvalues of the matrices
$\Lambda^m_{pp'}$ and ${\Lambda^1}^m_{pp'}$ in Eqs.
(\ref{eq:lambd-u_np-lambd}) and (\ref{eq:lambd-u_np-lambd-1}).  Let us
generalize the matrices $\Lambda_{pp'}^m$ and ${\Lambda^1}_{pp'}^m$
defined in Eqs.~(\ref{eq:lambd-=-frac2s-1}) and
(\ref{eq:lambd-=-frac4s}) to
\begin{align}
  \Lambda_{pp'}^m\Big(\frac{\sigma_{\perp}²}{N} \Big)=
   \frac{4\sigma_{\perp}² e^{\frac{-\sigma_{\perp}²}{N}(\gamma_p² +
       \gamma_{p'}²) } I_m \Big( \frac{2\sigma_{\perp}²}{N} \gamma_p
     \gamma_{p'} \Big)} {N {a_c}² J_m(X_{mp}) J_m(X_{mp'}) }, 
\end{align}
i.e. $\Lambda_{pp'}^m$ correspond to $N=2$ and ${\Lambda^1}_{pp'}^m$ correspond
to $N=1$.  One can then show that
\begin{align}\label{eq:sum_n-u_npm-u_np}
  \sum_n& U_{np}\mu_{mn}^M\lambda_{mn}^N U_{np'} = \notag \\ & \sum_s^M \left(
    \begin{array}{l}
      M\\s
    \end{array} \right) y^{M-s}(-4\sigma_{\perp}^2)^{-s} 
\Lambda_{pp'}^m\Big(\frac{\sigma_{\perp}²}{2(N+M-s)+s} \Big) 
\end{align}
This result along with the appropriate series expansion of functions
involving the eigenvalues $\lambda_{mn}$ and $\lambda^1_{mn}$ can be inserted into the
result for the propagator Eq. (\ref{eq:int-dk-sum_mn}), and the
resulting sum over indices $p$ and $p'$ takes the form
\begin{align}\label{eq:sum_pp-fracj_mg-e}
  \sum_{pp'} &\frac{J_m(\gamma_{p'}r')I_m(2\sigma_{\perp}²\gamma_px)
    e^{-\sigma_{\perp}² ( x²+\gamma_p² ) }}{a_c²J_{m+1}(X_{mp}) J_{m+1}(X_{mp'})}
  \Lambda_{pp'}^m\Big( \frac{\sigma_{\perp}²}{N} \Big)\notag  \\ &=
  \frac{1}{4\sigma_{\perp}²} e^{-
    \frac{{r'}^2}{4\sigma_{\perp}²} - \frac{N{r'}^2}{4\sigma_{\perp}²}
  } J_m( xr'),
\end{align}
where $N$ is an integer derived from Eq. (\ref{eq:sum_n-u_npm-u_np})
and the before mentioned series expansions. 
The propagator (\ref{eq:int-dk-sum_mn}) can therefore be written
\begin{widetext}
  \begin{align}
    P^{(+)}(\mathbf r,\mathbf r';t)&=\sum_{mn} \frac{ a |\mathcal
      D_{cl}| \kstokes^3\sqrt{\rho_0}(1-\partial_z^2)}{4\pi} \int xdx
    \int dy \frac{\sigma_{||}}{\sqrt{\pi}} e^{-\sigma_{||}^2y^2+iyz'}
    \sum_{pp'} U_{np} U_{np'}\times  \notag \\ & \hspace{2cm}
    4\sigma_{\perp}^2 e^{im\Delta \phi} \mathcal I_k \frac{
      J_m(xr)J_m(\gamma_{p'}r')I_m(2\sigma_{\perp}^2\gamma_px)}{a_c^2J_{m+1}(X_{mp})J_{m+1}(X_{mp'})}
    e^{-\sigma_{\perp}^2(\gamma_p^2+x^2)-\frac{{z'}^2}{4\sigma_{||}^2}}
    \notag \\ &= \frac{i a |\mathcal D_{cl}|\kstokes^3
    }{4\pi}\sqrt{\rho(\mathbf r)} \sum_m
    \int_0^{\sqrt{\frac{2}{\Delta z}}} xdx  \:e^{im\Delta\phi+i\Delta z} J_m(xr)J_m(xr')
   \times \notag \\ & \hspace{1cm} \sum_{l=0}^{\infty}
    \sum_{q=0}^{\infty} \left( \frac{i x^2 \Delta
        z}{2}\right)^l \left(\frac{i \lambda_0 t \Delta
        z^2}{\sqrt 8\sigma_{\perp^2}} \right)^q\: \Phi^q(r',z')\:
    \frac{I_{l+2q}\Big(2\sqrt{e^{-\frac{{r'}^2}{2\sigma_{\perp}^2}}
        \lambda_0 t \Delta
      z} \Big)}{\Big( \sqrt{e^{-\frac{{r'}^2}{2\sigma_{\perp}^2}} \lambda_0 t \Delta
          z}\Big)^{l+2q} }.
      \label{eq:-hspace6cm-big}
\intertext{where}
\Phi^q(r',z')&=\;\sum_{n=0}^q \sum_{s=0}^{E(n/2)}
\frac{e^{-\frac{{r'}^2}{4\sigma_{\perp}^2}(q+n)}}{(q-n)!(n-2s)!s!}
\left( \frac{-4i\sigma_{\perp}^2}{\sigma_{||}^2} \right)^n
(-\sigma_{||}^2)^s {z'}^{n-2s}
  \end{align}
\end{widetext}
We notice since $(x^2\Delta z)/2<1$, that choosing the variable
$\Delta z$ large means that the sum over $l$ will converge very fast.
Choosing the variable $\Delta z$ large can be done by placing the
detector plane far away from the sample, in which case we will talk
about a far-field calculation.  Unfortunately the sum over $q$
converges more slowly when $\Delta z$ is larger, and we can not quite
rely on our initial approximations [$\eta(k-k')\approx \delta(k-k')$,
see Sec.  \ref{sec:diagonalization}] for large $\Delta z$. We shall
therefore consider the problem in the near field region. The limit
$\sqrt{2/\Delta z}$ in the $x$-integral we shall on the other hand
approximate with the value $\sqrt{2/L}$, where
$L=\sqrt{2\pi}\sigma_{||}$ is the effective length of the atomic
ensemble. This approximation will become better at later times, since
the coherent build-up is essentially described by the modified Bessel
function $I_{l+2q}(2\sqrt{\lambda_0 \Delta z t})$ which in time will
dominate for large values of $\Delta z$. In Fig.
\ref{fig:draw-coher-build} we illustrate the physical significance of
the integral over $x$, which represents an integral over transverse
momentum.  We see that as we include more light from deviating angles,
this radiation has a shorter region over which it can build up, and as
the build-up is exponential in the build-up length, the error made by
the cut-off $L$ becomes relatively small.
From the propagator (\ref{eq:-hspace6cm-big}) the electric field
can be written, similar to the spontaneously emitted radiation,
(\ref{eq:d+_-mathbf-r-1}), as
\begin{align}\label{eq:d+_-mathbf-r-2}
  D^{(+)}_+&(\mathbf r,t) = D^{(+)}_+(\mathbf r,t)_0 + \int d^3r'\;
  P^{(+)}(\mathbf r,\mathbf r';t) \hat b^{\dagger}(\mathbf r',0).
\end{align}

\section{Intensity and the correlation
  function}\label{sec:intens-corr-funct}

In this section we consider the electric field, and assume that we
place a detector in a plane at some position $z_0$ after the end of
the atomic sample. We then define the correlation function as a
function of the radial coordinate $r$ and time $t$
\begin{align}
  \mathcal C(r,r',t)=\hspace{-2pt}\frac{2}{\hbar \epsilon_0
    \kstokes}\hspace{-2pt}\int \hspace{-2pt} d\phi \<
  \hat D^{(-)}_+(z_0,r,\phi,t) \hat D^{(+)}_-(z_0,r',\phi,t) \>,
\end{align}
where $\< \cdot \>$ is the quantum mechanical average. The
normalization $\frac{2}{\hbar \epsilon_0 \kstokes}$ is chosen such
that the number of photons in a pulse is given by 
\begin{align}
  N_P=\int  \frac{dA}{\kstokes^2} \int dt \mathcal C(r,r,t).
\end{align}
The factor $\kstokes^2$ is inserted since lengths are measured in
units of $\kstokes$.  Inserting the propagator in Eq.
(\ref{eq:-hspace6cm-big}) allows us to describe SRS, while the
propagator (\ref{eq:sum_mn-frac-mathcal}) gives the spontaneous
emission for short times. We shall be most interested in SRS, but will
also for comparison examine the spontaneously emitted light. First we
present the correlation function describing the SRS, when measured in
a plane at the end of the atomic sample.  An important parameter below
will be the Fresnel number $\mathcal F$ which we define by $\mathcal
F=\frac{\sigma_{\perp}^2}{L}$. [Recall that all lengths are measured
in units of $\kstokes$.] We shall in general assume the Fresnel number
to be large, in particular $\mathcal F>1$. In the integration over
$z'$ we will use the following substitution
\begin{align}\label{eq:int-dz-efracz22s}
  \int dz' e^{\frac{z'^2}{2\sigma_{||}^2}} \rightarrow \int_0^L dz',
\end{align}
where $L=\sqrt{2\pi}\sigma_{||}$.
The correlation function can then be calculated to give 
\begin{widetext}
  \begin{align}\label{eq:mathcal-cr-r}
    \mathcal C(r,r',t)=& \frac{\kstokes^2 \lambda_0 e^{-\Gamma t}}{4
      \mathcal F} \sum_m \sum_{\substack{lqk\\l'q'k'}}
    \sum_{n,n'}^{q,q'} \hspace{-2pt}\int_0^{2\mathcal
      F}\hspace{-12pt}dy \hspace{-2pt}\int_0^{2\mathcal
      F}\hspace{-12pt} dy' \Bigg\{ \Big( \frac{-i y}{2\mathcal
      F}\Big)^l \Big(\frac{i y'}{2\mathcal F} \Big)^{l'}
    \Big(\frac{-i}{\sqrt 8 \mathcal F}\Big)^q \Big(\frac{i }{\sqrt 8
      \mathcal F}\Big)^{q'} (8i\pi\mathcal
    F)^n (-8i\pi\mathcal F)^{n'} \times \notag \\
    & \hspace{3cm} J_m\big(\sqrt{y}\frac{r}{\sigma_{\perp}}\big)
    J_m\big(\sqrt{y'} \frac{r'}{\sigma_{\perp}}\big)
    e^{-\frac{y+y'}{2+2(k+k')+q+q'+n+n'}}
    I_m\Big({\scriptstyle{\frac{2\sqrt{yy'}}{2+2(k+k')+q+q'+n+n'}}}\Big)\times
    \notag \\ & \hspace{7cm} \chi^{l'q'k'n'}_{lqkn} \frac{ (\lambda_0
      t L)^{k+k'+q+q'} }{k!k'!(l+2q+k)!(l'+2q'+k')!} \Bigg\},
    \intertext{where} &\chi^{l'q'k'n'}_{lqkn}=
    \sum_{s,s'}^{\substack{E(n/2),\\E(n'/2)}}
    \sum_{Q,Q'}^{\substack{n-2s,\\n'-2s'}} \frac{2\left(
        \frac{(-1)^{Q+Q'+s+s'} (2 \pi)^{-s-s'}}{(q-n)!
          (q'-n')!(n-2s-Q)!(n'-2s'-Q')!s!s'!  Q!Q'!}\right)
    }{\scriptstyle{(1+Q+Q'+k+k'+l+l'+2(q+q'))(2+2(k+k')+q+q'+n+n')}}.
  \end{align}
\end{widetext}
This is the main result of this section. We notice that when $r$ is
measured in units of $\sigma_{\perp}$, the only
variables controlling the behavior of the correlation function is the
Fresnel number, $\mathcal F$, the optical depth, $d=6 \pi \rho_0 L$ and time
measured in units of the single atom scattering rate $\Gamma$. This
follows since $\lambda_0 t L= \frac{3\pi}{2} \rho_0 L \Gamma t
=\frac{d\Gamma t}{4}$. From
the correlation function (\ref{eq:mathcal-cr-r}) we also expect
fast convergence in the index $q$ and $l$ as the Fresnel number
increases.  In the remainder of this article we shall evaluate the
correlation function numerically. Even though the correlation function
involves a double integral beside the large number of sums, we see
that as we increase the index $k,k',q,q',n,n'$, the $y$- and
$y'$-integrals will simplify. This follows since the argument of the
modified Bessel function decreases as the indices $k,k',q,q',n,n$
increases. We can therefore use the small argument limit. Similarly the
Gaussian function can be approximated by unity.  From Eq.
(\ref{eq:mathcal-cr-r}) we see that the dominating term in the
sum over $k$ will have a higher $k$ when time grows. This means that
the radial behavior of the beam simplifies. Due to the small
argument description of the modified Bessel function the radiation is eventually
dominated by the $m=0$ mode.

\subsection{Intensity on the symmetry axis}\label{sec:intens-symm-axis}

In this section we will examine the radiated light on the
symmetry axis. The purpose is to examine the timescale on which there
is a crossover from spontaneous emission to SRS.

Placing the detector on the symmetry axis is a nice simplification
especially for the spontaneous emission correlation function, since in
that case we may use the result presented in Eqs.
(\ref{eq:d+_-mathbf-r-1}) and (\ref{eq:-=frac-mathcal}).  Also the
SRS correlation function simplifies since terms with
$m\neq 0$ vanish at the symmetry axis.  In the spontaneous emission
limit $t\approx 0$ the intensity on the axis is given by
\begin{align}
  \mathcal C_0(0,0)= \kstokes^2 \lambda_0 \int_0^L& d\Delta z \int r'dr' \:
  e^{-\frac{{r'}^2}{2\sigma_{\perp}^2}}
  \frac{(\frac{1}{2}{r'}²+\Delta z²)²}{({r'}²+\Delta z²)³},
\end{align}
where we use the substitution in Eq. (\ref{eq:int-dz-efracz22s}), and
assume that the detector is placed at the end of the atomic ensemble. 
The $z$-integral can be performed analytically and one finds
\begin{align}\label{eq:mathc-c_00-0=leftfr}
  \mathcal C_0(0,0)=& \kstokes ^2 \lambda_0 L \int r dr
  \:\frac{e^{-\frac{L^2}{2\sigma_{\perp}^2}{r}^2}}{32} \times \notag
  \\ &
 \Bigg\{  \frac{-13-11r^2}{(1+r^2)^2}+ \frac{19 \arctan
    ( r^{-1} )}{r} \Bigg\}.
\end{align}
From this expression we find that the parameters
controlling the intensity on the symmetry axis is the optical
depth, and the ratio between the length and the width of the atomic
ensemble. 

\begin{figure}
  \centering
  \includegraphics[width=0.48\textwidth]{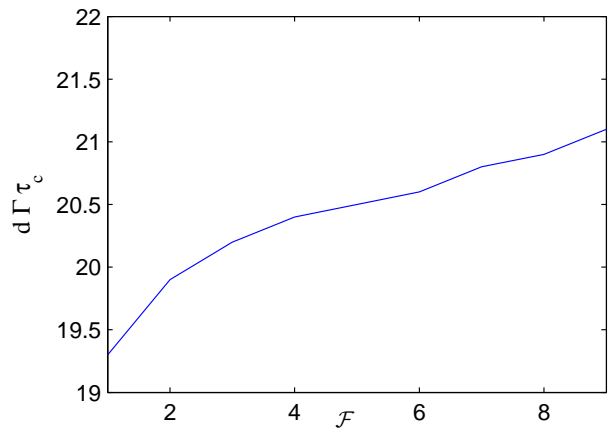}
  \caption{Plot of the time $\tau_{c}$ measured in units of $(d
    \Gamma)^{-1} $, at which the intensity on the symmetry axis is
    dominated by SRS.  The cross-over time is only weakly dependent on
    Fresnel number, and is given primarily by the optical depth.}
  \label{fig:plot-time-logar}
\end{figure}

We shall now investigate the time scale on which SRS begins to
dominate the radiation. For short times where the radiation is
dominated by spontaneous emission, we expect that the radiation is
being emitted almost homogeneously in all directions, so that it makes
sence to compare the spontaneous emission in a given direction, with
SRS. We find from Eq. (\ref{eq:mathc-c_00-0=leftfr}) that the figure
of merit for the spontaneous emission is the density, the length, and
the width of the atomic ensemble, and not as in the case of SRS, only
the Fresnel number and the optical depth.  Thus in order to compare
the two time domains, the spontaneous emission and the SRS, we will
have to fix e.g. the length of the system. From Eq.
(\ref{eq:mathcal-cr-r}) we find that the cross-over time when going
from spontaneous emission to SRS scales linearly with the optical
depth, so that an increase of the optical depth gives a similar
decrease of the cross-over time. In Fig.  \ref{fig:plot-time-logar} we
show this cross-over for varying Fresnel numbers $\mathcal F$ and a
fixed length of the ensemble $L=300\frac{\lambda_s}{2\pi}$.  We see
that the cross-over only depends weakly on the Fresnel number. The
main parameter characterizing the time scale is thus the optical
depth. The cross-over time is found by plotting the intensity on the
symmetry axis, Eq. (\ref{eq:mathcal-cr-r}) and the spontaneous
emission on the symmetry axis, Eq. (\ref{eq:mathc-c_00-0=leftfr}), and
finding the point at which they cross.

\subsection{Intensity profile} \label{sec:intensity}

In this section we shall look at the spatial shape of the 
radiation leaving the atomic ensemble.  Before we present
the numerical calculations for the coherent emission we will look at
the correlation function in Eq. (\ref{eq:mathcal-cr-r}). The spatial
shape of the function is mainly given by 
\begin{align}\label{eq:int_0k-1dy-int_0k}
  \int_0^{2 \mathcal F}&dy \int_0^{2 \mathcal F} dy'
  J_m\big(\sqrt{y}\frac{r}{\sigma_{\perp}}\big) J_m\big(\sqrt{y'}
  \frac{r'}{\sigma_{\perp}}\big)\times \notag \\ &e^{-\frac{y+y'}{2+2(k+k')+q+q'+n+n')}}
  I_m\Big(\scriptstyle{\frac{2\sqrt{yy'}}{2+2(k+k')+q+q'+n+n'}}\Big)
\end{align}
With increasing values of $k,k',q,q',n$ and $n'$, the exponential
function can to a higher and higher precision be approximated by
unity. The modified Bessel function of order $m$ can for small
arguments be approximated with an $m$'th order polynomial
\begin{align}\label{eq:i_mz-appr-fracz2mm}
  I_m(z) \approx \frac{(z/2)^m}{m!}.
\end{align}
From the argument of the modified Bessel function in Eq.
(\ref{eq:int_0k-1dy-int_0k}) we find that the region for which the
approximation Eq. (\ref{eq:i_mz-appr-fracz2mm}) is applicable is given
both by the number $2+2(k+k')+q+q'+n+n'$ and by the integration range
$2\mathcal F$. Eq.  (\ref{eq:int_0k-1dy-int_0k}) indicates that as time
increases the dominant mode will be the $m=0$ mode for a finite sized
atomic ensemble. On the other hand we see that for an infinitely sized
atomic ensemble all $m$-modes will contribute. This is essentially the
limit considered in the one-dimensional theory in Ref.
\cite{raymer-mostow}.  That theory applies to an infinitely wide
sample such that all modes experience the same dynamics. For a sample
of finite width we see that the oscillating behavior of the Bessel
functions $J_m$ gives a cut of the width of the beam scaling with
approximately $r_c/\sigma_{\perp} \sim 1/\sqrt{2 \mathcal F}$ or $r_c
\sim \sqrt{L/2}$.  This cut $r_c$ will, due to the behavior of the
Bessel function $J_m$, increase as $m$ increases.  We thus see that
even though the width of the beam is mainly determined by the length
of the atomic ensemble, the width of the atomic ensemble plays an
important role as a wider ensemble supports higher order modes that
are inherently wider, thus in effect a wider atomic ensemble will
generate a wider beam.

\begin{figure}
  \centering
  \vspace{5pt}
  \includegraphics[width=0.48\textwidth]{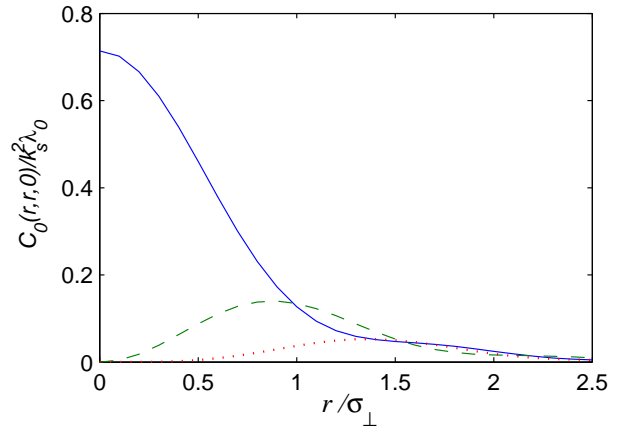}\\
  \caption{(Color online) Plot of the radiated power for different
    azimuthal quantum numbers $m=0,\pm 1,\pm 2$ as a function of the
    detection coordinate $r/\sigma_{\perp}$. The plot is taken at the
    initial time, $\Gamma t=0$ with an optical depth $d=160$.
    Comparison with Fig. \ref{fig:sigma-20-time-radial222} demonstrate
    how the relative distribution of radiation with different
    azimuthal quantum number $m$ is changed as the Fresnel number
    $\mathcal F$ is varied. Here we use $\mathcal F=4$ and in Fig.
    \ref{fig:sigma-20-time-radial222}) we use $\mathcal F=8$. The
    solid line correspond to $m=0$, the dashed line correspond to
    $m=\pm 1$ and finally the dotted line correspond to $m=\pm 2$.}
  \label{fig:sigma-20-time-radial}
\end{figure}

\begin{figure}
  \centering
  \vspace{5pt}
  \includegraphics[width=0.48\textwidth]{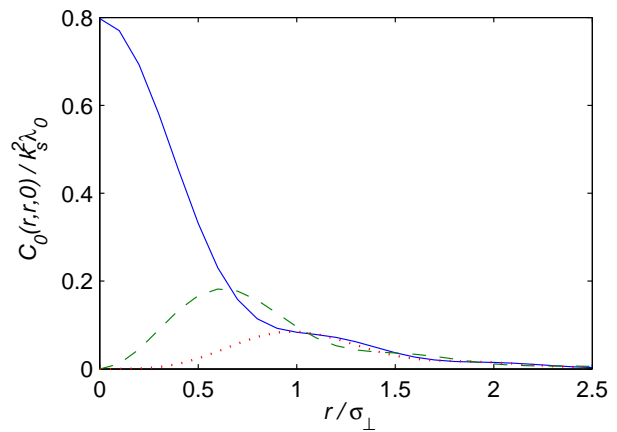}
  \caption{(Color online) Same as Fig. \ref{fig:sigma-20-time-radial} but with
    Fresnel number $\mathcal F=8$ }
  \label{fig:sigma-20-time-radial222}
\end{figure}

From the expansion Eq.  (\ref{eq:mathcal-cr-r}) and the small argument
limit of the modified Bessel function Eq.
(\ref{eq:i_mz-appr-fracz2mm}) along with
Eq.~(\ref{eq:int_0k-1dy-int_0k}) we see that as time increases the
contributions to the intensity from modes $m\neq 0$ will not grow as
rapidly as $m=0$.  In Fig.  \ref{fig:sigma-20-time-radial} we show a
plot of the radiated power in three SRS modes at time $t=0$, where we
use a Fresnel number $\mathcal F=4$ and optical depth $d=160$. In Fig.
\ref{fig:sigma-20-time-radial222} we use an atomic ensemble with
Fresnel number $\mathcal F=8$ and optical depth $d=160$.  The plots
demonstrates how the relative importance between different modes are
changed as the Fresnel number is changed. From the two plots in Figs.
\ref{fig:sigma-20-time-radial} and \ref{fig:sigma-20-time-radial222}
we see that the larger the Fresnel number, the more modes with higher
azimuthal quantum number $m$ can we fit into the system. In Fig.
\ref{fig:sigma-20-time-radial} we see that the principal mode $m=0$ is
dominating the higher order modes.  When the Fresnel number is doubled
in Fig.  \ref{fig:sigma-20-time-radial222} the principal mode $m=0$ is
still dominating, but less than in Fig.
\ref{fig:sigma-20-time-radial}.  To conclude that a higher Fresnel
number, allows higher order azimuthal quantum numbers $m$ to
contribute, we have to look at the total number of photons for each
$m$. This is the topic of Sec.  \ref{sec:total-coher-radi}, and from
the results derived there we indeed find that we can have relatively
more photons for higher order $m$ as the Fresnel number is increased.
E.g for $\mathcal F=4$ the photon power in each of the $m=\pm 1$ mode
relative to the $m=0$ mode is about 62\%, and it is 35\% for $m=\pm
2$, whereas for $\mathcal F=8$ this number is increased to 72\% for
$m=\pm 1$ and 49\% for $m=\pm 2$.

\begin{figure}
  \centering
  \vspace{5pt}
  \includegraphics[width=0.48\textwidth]{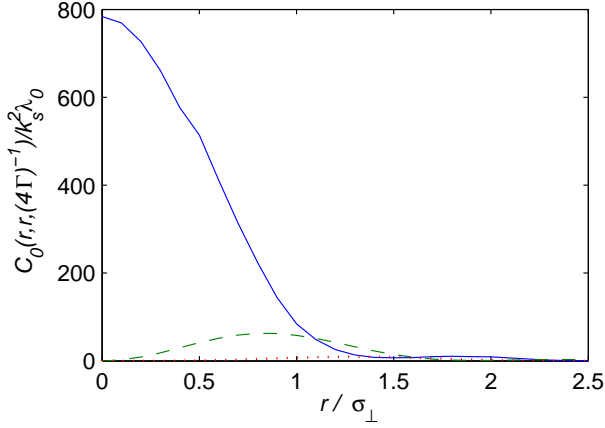}\\
  \caption{(Color online) Plot of the radiated power for different
    azimuthal quantum numbers $m=0$ (solid line), $m=\pm 1$ (dashed
    line), and $m=\pm 2$ (dotted line) as a function of the detection
    coordinate $r/\sigma_{\perp}$. Here the plot is made for a time of
    $\Gamma t=0.25$ and an optical depth $d=160$. Comparison with Fig.
    \ref{fig:see-caption-fig} demonstrate how the relative
    distribution of radiation in modes with different $m$ is changed as the
    Fresnel number $\mathcal F$ is varied. Here we use $\mathcal F=4$
    and in Fig. \ref{fig:see-caption-fig} we use $\mathcal F=8$.  When
    the plots in Figs. \ref{fig:color-online-plot} and
    \ref{fig:see-caption-fig} are compared with the plots for $t=0$ in
    Figs. \ref{fig:sigma-20-time-radial} and
    \ref{fig:sigma-20-time-radial222}, we indeed see that as time
    increases, the evolution of the principal mode, $m=0$ is faster
    than that of the higher order modes. }
  \label{fig:color-online-plot}
\end{figure}

\begin{figure}
  \centering
  \vspace{5pt}
  \includegraphics[width=0.48\textwidth]{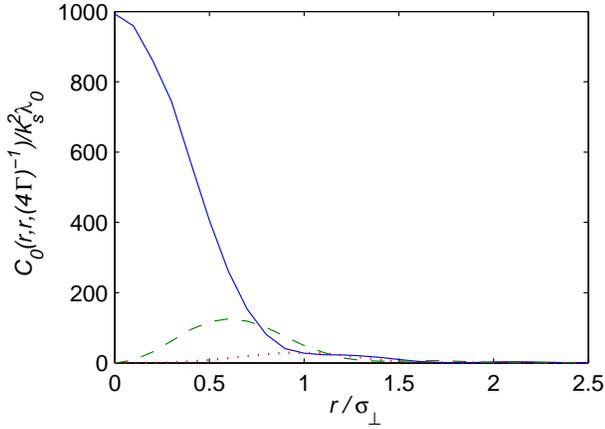}
  \caption{(Color online) Same as Fig. \ref{fig:color-online-plot},
    but with Fresnel number $\mathcal F=8$ }
  \label{fig:see-caption-fig}
\end{figure}

Next we consider how the time evolution changes the shape of the mode.
From the earlier discussion of Eq. (\ref{eq:int_0k-1dy-int_0k}) we
expect that the relative photon intensity carried by modes with $m$
different from the principal mode $m=0$ will decrease compared to the
principal mode as time is increased. In Figs.
\ref{fig:color-online-plot} and \ref{fig:see-caption-fig} we plot the
radial distribution of the photon power at time $\Gamma t=0.25$. We
see that the radial shape of the modes have not changed compared with
the plots at $t=0$, [ Figs. \ref{fig:sigma-20-time-radial} and
\ref{fig:sigma-20-time-radial222} ].  The relative maximum photon
power for modes with $m\neq 0$ has however decreased compared with the
principal mode $m=0$. Again we can look at the total photon power in
each mode, and find that for the case of Fresnel number $\mathcal
F=4$, each of the modes $m=\pm 1$ now only contains 23\% of the
intensity carried by the $m=0$ mode, and the $m=\pm 2$ mode only
4.8\%. A similar behavior is found for the $\mathcal F=8$ case, though
less pronounced, i.e. now each of the modes $m=\pm 1$ carries 38\% of the
photon power compared with the $m=0$ mode, and for the $m=\pm 2$ modes
it is 12\%. As expected the modes with $m\neq 0$ become relatively
less important for long times.

\noindent

\noindent

\subsection{Total coherent radiation.}\label{sec:total-coher-radi}

\begin{figure}
  \centering
  \includegraphics[width=0.48\textwidth]{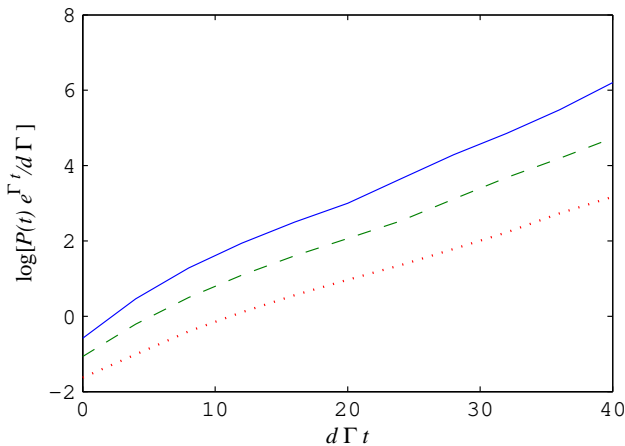}
  \caption{(Color online) Plot of the total radiated power $P$
    measured in number of photons, $N_p$ per $d \Gamma $. We in
    addition scale out the natural decay $e^{-\Gamma t}$. With this
    scaling we get a universal curve applying to all ensembles with
    the same Fresnel number. The time axis is scaled in units of
    $d\Gamma$. We use a Fresnel number of $\mathcal F=4$ and show
    results for three different $m$-modes, $m=0$ (solid line),
    $m=\pm 1$ (dashed line), and $m=\pm 2$ (dotted line).  We see that the
    principal mode $m=0$ has a slightly faster growth than higher
    order modes.}
  \label{fig:plot-total-intensity}
\end{figure}

Finally we will examine the total intensity of
SRS. We shall in this section not only show
the effect of the analytical calculations made so far, but also compare
the result with a purely numerical treatment of the equations given
in Eq.  (\ref{eq:fracddt-hat-bdagg-1}). The total intensity is
normalized such that it gives the number of photons per second coming
through the detector-plane 
\begin{align}\label{eq:i_tt=fr-epsil-hbar}
  P(t)=\frac{2}{\kstokes \epsilon_0 \hbar}\int &\frac{rdr}{\klaser^2}
  \int d\phi  \times \notag \\ &\< \hat D^{(-)}_+(z_0,r,\phi,t) \hat
  D^{(+)}_-(z_0,r',\phi,t) \>.
\end{align}
To find the total intensity we use the result in
Eq.(\ref{eq:mathcal-cr-r}) and perform the radial integral. To do this we
use the relation
\begin{align}\label{eq:int_0-rdr-j_mxrj_mxr}
  \int_0^\infty rdr J_m(xr)J_m(x'r)=\frac{\delta(x-x')}{x},
\end{align}
derived in Appendix \ref{sec:compl-sec.-total}. The total
radiation is then found to be
\begin{widetext}
  \begin{align}\label{eq:i_tt=fr-sum_s}
    P(t)=&  \frac{d \Gamma  e^{-\Gamma t}}{8}
    \sum_m \sum_{\substack{lqk\\l'q'k'}} \sum_{n,n'}^{q,q'}
    \int_0^{2 \mathcal F}dy  \Bigg\{
      \Big( \frac{-i y}{2\mathcal F}\Big)^l
    \Big(\frac{i y}{2\mathcal F} \Big)^{l'} \Big(\frac{-i}{\sqrt 8 \mathcal
      F}\Big)^q \Big(\frac{i }{\sqrt 8 \mathcal F}\Big)^{q'} (8i\pi\mathcal
    F)^n (-8i\pi\mathcal F)^{n'} \times \notag \\ 
    &\chi^{l'q'k'n'}_{lqkn} \frac{ (4d\Gamma t)^{k+k'+q+q'} }{k!k'!(l+2q+k)!(l'+2q'+k')!}
     e^{-\frac{2y}{2+2(k+k')+q+q'+n+n'}}
    I_m\Big(\scriptstyle{\frac{2y}{2+2(k+k')+q+q'+n+n'}}\Big) \Bigg\}.
   \end{align}
\end{widetext}
In Fig. \ref{fig:plot-total-intensity} we show a plot of the total
radiated power, Eq. (\ref{eq:i_tt=fr-sum_s}) for the parameter
$\mathcal F=4$. The scaling is chosen such that the curve will
be identical for all samples with the same Fresnel number $\mathcal
F$.  It is interesting to note that indeed the intensity in modes with
$m\neq 0$ evolves slower in time than for the $m=0$ mode.  This can be
seen by looking at the slope of the curves as they are plotted on a
logarithmic scale.

\begin{figure}
  \centering
  \includegraphics[width=0.48\textwidth]{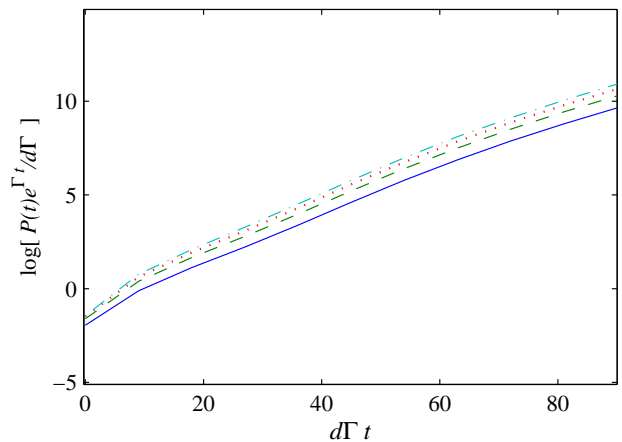}
  \caption{(Color online) Plot of the total radiated power calculated
    for varying Fresnel numbers. The lines are calculated using the
    expression Eq. (\ref{eq:i_tt=fr-sum_s}) for the principal mode
    $m=0$. Apart from a complicated behavior at short times we see
    that the total radiation is linearly proportional to the Fresnel
    number. This can also be seen from
    Eq. (\ref{eq:i_tt=fr-sum_s}). The solid line correspond to
    $\mathcal F=1$, the dashed line to $\mathcal F=2$, the dotted
    line to $\mathcal F=3$ and the dash-dotted curve correspond to
    $\mathcal F=4$.}
  \label{fig:plot-total-radiated}
\end{figure}

In Fig. \ref{fig:plot-total-radiated} we analyze how the total
radiation depends on the Fresnel number. For large times, the
dependence is approximately linear in the Fresnel number. This may
also be concluded directly from Eq.  (\ref{eq:i_tt=fr-sum_s}).

We now compare the result for the total radiated power with the
effective one-dimensional calculation derived in Ref.
\cite{raymer-mostow}. The general assumption in the one-dimensional
calculation is that the atomic ensemble is infinitely wide. This
assumption makes the problem easy to solve in Fourier space. When
the transverse momentum in the propagator for the light is 
neglected, the result for the total radiated power is that all modes
corresponding to different transverse momentum gives equal
contribution to the total radiated power. Thus the total radiated
power measured in units of number of photons per time gives
\begin{align}\label{eq:i_trmt=-sum_m-k_perp}
  P^{RM}(t)= \sum_{\mathbf k_{\perp}} \frac{d \Gamma e^{-\Gamma t}}{4}\Big(
  I^2_0(\sqrt{d\Gamma t}) - I^2_1(\sqrt{d\Gamma t}) \Big).
\end{align}
Since we have neglected all information on the transverse shape there
is a priori no upper limit on the transverse momentum. Thus taking all
modes corresponding to all transverse momentum into account gives an
infinite contribution. A derivation of such a mode description can be
found in Ref.  \cite{hammerer}. It is concluded in Ref.
\cite{raymer-mostow2} that for a Fresnel number near unity the
radiation is dominated by a single transverse mode, and thus the total
radiation is finite, and given approximately by a single term in the
sum (\ref{eq:i_trmt=-sum_m-k_perp}).  

We can also make a simplification of our result
(\ref{eq:i_tt=fr-sum_s}) by neglecting all kinds of finite size
effects in the eigenvalue matrix, $M^{kmn}_{k'm'n'}$. From the
derivation of Eq. (\ref{eq:i_tt=fr-sum_s}), one sees that this amounts
to fixing $\{q,q',l,l'\}=0$ and setting $k=0$ and $k'=0$ 
in the modified Bessel function as well as the exponential function.
Finally the approximation gives an additional factor of $1+k+k'$.
This is an oversimplification, but allows a comparison with the
results by Raymer and Mostowski in Ref.  \cite{raymer-mostow}.  The 
total radiated power is then given by
\begin{align}
  P_0(t)=\frac{d \Gamma e^{-\Gamma t}}{4}\Big(
  I^2_0&(\sqrt{d \Gamma t}) - I^2_1(\sqrt{d \Gamma t})
  \Big)\times \notag \\ &\int_0^{2\mathcal F} dy \sum_m \frac{e^{-y} I_m(y)}{2} 
\end{align}
For $\mathcal F \approx 1$ this expression is identical to a single
term in the sum in Eq. (\ref{eq:i_trmt=-sum_m-k_perp}).  We now assume
the Fresnel number $\mathcal F \sim 1$, and apply the approximation
(\ref{eq:i_mz-appr-fracz2mm}), which is only valid for small Fresnel
numbers. In this way we find
\begin{align}
  \sum_m \frac{e^{-y} I_m(y)}{2} \approx e^{-\frac{y}{2}} - \frac{e^{-y}}{2},
\end{align}
and the integral results in the total radiated power
\begin{align}\label{eq:i_tt=lambda_0-l-e}
  P_0(t)=\frac{d \Gamma e^{-\Gamma t}}{4}(\frac{3}{2}
  \hspace{1pt}-\hspace{2pt} & 2 e^{\footnotesize{-\mathcal F}}+
  \frac{e^{\footnotesize{-2\mathcal F}}}{2})\times \notag \\ &\Big(
  I^2_0(\sqrt{d \Gamma t}) - I^2_1(\sqrt{d \Gamma t}) \Big).
\end{align}
 We are thus led to conclude that for
a Fresnel number near unity, the simple Raymer Mostowski result
correspond to neglecting all spatial corrections to the dynamic of the
atoms and also neglecting spatial corrections to the propagation of
light out of the atomic ensemble. 

We can improve the approximation, by looking at the general result in
Eq. (\ref{eq:i_tt=fr-sum_s}) and keeping only zeroth order terms in
the index $q,q',l$ and $l'$. In this way we get
\begin{align}\label{eq:pt=-fracd-gamma}
      P_1(t)=  \frac{d \Gamma  e^{-\Gamma t}}{8}
    \sum_m \sum_{\substack{kk'}} &
    \int_0^{2 \mathcal F}dy  \Bigg\{ e^{-\frac{y}{1+k+k'}}
    I_m\Big(\scriptstyle{\frac{y}{1+k+k'}}\Big)
       \times \notag \\ 
    &\frac{ (d\Gamma t/4)^{k+k'} }{k!^2k'!^2(1+k+k')^2}
     \Bigg\}.
\end{align}
for $\mathcal F \ll \frac{1}{2} + \frac{d \Gamma t}{8}$ we can reduce
Eq. (\ref{eq:pt=-fracd-gamma}) even further and arrive at the result
\begin{align}
  P_1(t)\approx \frac{\mathcal F d \Gamma e^{-\Gamma t}}{4} \Big(
  I_0^2(\sqrt{d\Gamma t}) &- 2 I_1^2(\sqrt{d\Gamma t}) \notag \\ & +
  I_0(\sqrt{d\Gamma t})I_2(\sqrt{d\Gamma t}) \Big). 
\end{align}
In this limit $\mathcal F \ll \frac{1}{2} + \frac{d \Gamma t}{8}$ the
only contribution to the total radiated power comes from the $m=0$ mode.

\begin{figure}
  \centering
  \includegraphics[width=.48\textwidth]{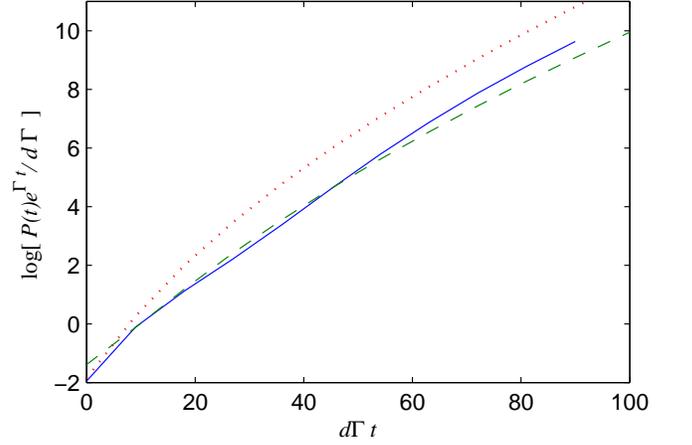}
  \caption{(Color online) Plot of the total radiated power $P(t)$
    scaled so that it only depend on Fresnel number. Here we use
    $\mathcal F=1$. To demonstrate the effects of a finite sized atomic
    ensemble, we show three different curves. The solid line is $P(t)$,
    the general result from Eq. (\ref{eq:i_tt=fr-sum_s}). The
    dashed line is $P_1(t)$ (\ref{eq:pt=-fracd-gamma}), where we use the
    zeroth order expansion of the general result
    (\ref{eq:i_tt=fr-sum_s}) and assume a large value of $d\Gamma t$.
    Finally the dotted line is $P_0(t)$ (\ref{eq:i_tt=lambda_0-l-e}),
    where we completely neglect all geometric effects on the
    matrix $M^{kmn}_{k'm'n'}$.}
  \label{fig:plot-total-radiated-1}
\end{figure}

In Fig. \ref{fig:plot-total-radiated-1} we analyze how the different
corrections to the Raymer Mostowski calculation effects the total
radiated power. We fix the Fresnel number at $\mathcal F = 1$, as this
is the limit where the Raymer Mostowski result is assumed to be valid.
The curve $P_0(t)$ is the simple result Eq.
(\ref{eq:i_tt=lambda_0-l-e}). In curve $P_1(t)$ we use the lowest
order finite size correction, that is Eq.  (\ref{eq:pt=-fracd-gamma}).
Finally in curve $P(t)$ we use the general result from Eq.
(\ref{eq:i_tt=fr-sum_s}), which is evaluated nummerically with the
approximated Bessel function (\ref{eq:i_mz-appr-fracz2mm}). The
approximation is used so that we gan get an estimation of the effect
of all azimuthal quantum numbers, thus the nummerical methods require
modest Fresnel numbers, such that the main contribution to the total
radiated power comes from the mode coresponding to $m=0$. We see that
the simple Raymer Mostowski type result, Eq.
(\ref{eq:i_tt=lambda_0-l-e}) over-estimates the total radiated power
compared to the general result. We also see that the zeroth order
result $P_1(t)$ is a much better approximation in the regime $d\Gamma
t/8 \gg \mathcal F$.

Finally we compare the result of Eq. (\ref{eq:i_tt=fr-sum_s}) with a
purely numerical calculation based on the point particle equations
(\ref{eq:fracddt-hat-bdagg-1}) and (\ref{eq:mathbf-d+mathbf-r}).
To make such a comparison we need to connect the evolution of the
atomic operators $\hat b_j(t)$ with the total intensity of the
radiated field. Based on energy conservation, the evolution
of the number of atoms in the ground state, is given by
the number of photons exiting a boundary sphere enclosing the atomic
ensemble. We derive this conservation law in Appendix
\ref{sec:sum-rule} where we show that 
\begin{align}\label{eq:frac2kl-hbar-epsil}
  \frac{2}{\kstokes \hbar \epsilon_0} \int d\Omega \;& \mathbf D^{(-)}(\mathbf
  r,t) \cdot \mathbf D^{(+)}(\mathbf r ,t)\notag \\ & =\sum_{jj'} \Big\{ \tilde
  M_{jj'} \hat b_j(t) \hat b_{j'}^{\dagger}(t) + H.c. \Big\}, 
\end{align}
where $\tilde M_{jj'}$ is given by $M_{jj'}+\Gamma \delta_{jj'}$, and
$M_{jj'}$ is given in Eq. (\ref{eq:m_jj-=-frac}). When comparing the
result of Eq. (\ref{eq:i_tt=fr-sum_s}) to the atomic evolution we have
to remember that we are only measuring half of the photons, since we
only consider the emission at one end of the ensemble. Using that the
evolution of the atomic operators are given by
\begin{align}
  \frac{d}{dt} b_j^{\dagger}(t) = \sum_{j'} M_{jj'} b^{\dagger}_{j'}(t),
\end{align}
we find that the atomic operators evolve in time according to
\begin{align}
  b^{\dagger}_j(t)=\sum_{j'}e^{\bar{\bar M}t}|_{jj'} b^{\dagger}_{j'},
\end{align}
where we define $\bar{\bar M}$ as the matrix with elements given by
$M_{jj'}$. After taking quantum average of the result in Eq.
(\ref{eq:frac2kl-hbar-epsil}) we find that
\begin{align}
  \frac{2}{\kstokes \hbar \epsilon_0} \int d\Omega \;& \< \hat
  D^{(-)}_+(\mathbf r,t) \hat
  D^{(+)}_-(\mathbf r,t) \>\notag \\ & = \text{trace}\big[
  {e^{\bar{\bar M}^*t}}^T (\openone + \bar{\bar M})e^{\bar{\bar M} t}
  \big] + C.c.
\end{align}
We then find the total intensity from the point particle model
\begin{align}\label{eq:ipp_tt=fr-sum_n-big}
  P_N(t)=\frac{1}{2}\Big\{\text{trace}\big[
  {e^{\bar{\bar M}^*t}}^T (\openone + \bar{\bar M})e^{\bar{\bar M} t}
  \big] + C.c.\Big\},
\end{align}
where we normalize with a factor $1/2$ since we want to compare the
result with the result in Eq. (\ref{eq:i_tt=fr-sum_s}). 

The advantages of making these calculations, or indeed solving the
problem of SRS on a computer are clear. One avoids the
problems of shifting from the point particle model to a continuous
model. Thereby one also automatically include dipole dipole
interaction effects connected to the point particle nature of the
system which we have ignored here. Also the computer easily
describes the total radiated field and not only the strongest
super-radiating mode as we have analyzed here. On the other hand the
direct method is numerically heavy for a large number of atoms, and we
are limited to $N\sim 6000$ atoms. To understand the behavior at
larger number of atoms it is therefore important to have an analytical
theory along the lines considered here.

\begin{figure}
  \centering
  \includegraphics[width=0.48\textwidth]{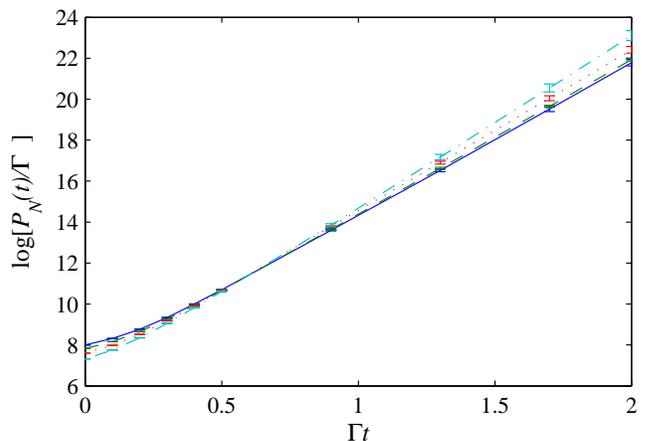}
  \caption{(Color online) Nummerical results for the total radiated
    power $P_N(t)$ per decay rate $\Gamma$ in the point particle model
    (\ref{eq:ipp_tt=fr-sum_n-big}). To exploit the nummerical model we
    fix the Fresnel number $\mathcal F=4$ and the optical depth
    $d=90$, but vary the number of atoms involved. The solid line
    correspond to $N=6000$ atoms, the dashed line to $N=5000$, the dotted line
    to $N=4000$, and finally the dash-dotted line correspond to
    $N=3000$ atoms. As the plot shows there is a dependence on the
    atomic density due to point particle effects that is not included
    in the analytical theory, but as the atomic density decreases
    (with increasing $N$) the curves seem to converge. The errorbars
    indicate the noise inherent in the point particle model due to the
    random positions of the atoms.}
  \label{fig:plot-show-comp}
\end{figure}

\begin{figure}
  \centering
  \includegraphics[width=0.48\textwidth]{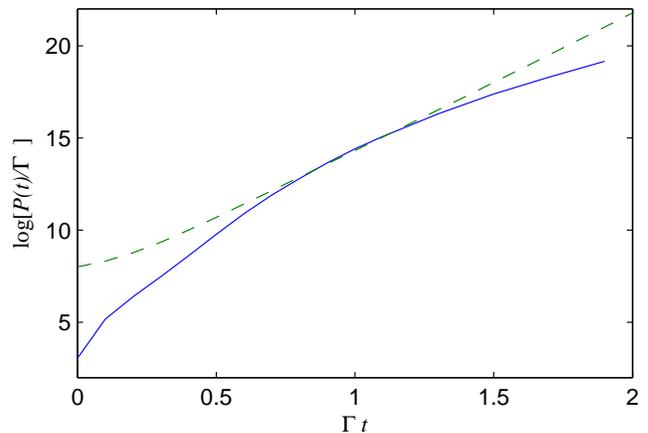}
  \caption{(Color online) Comparison of the analytical calculation of
    the radiated power $P(t)$, (\ref{eq:i_tt=fr-sum_s}), (solid
    line), with the nummerical result $P_N(t)$,
    (\ref{eq:ipp_tt=fr-sum_n-big}), (dashed line). The Fresnel number
    is $\mathcal F=4$ and the optical depth is $d=90$. }
\label{fig:color-online-here}
\end{figure}

To make the numerical simulation we have randomly distributed between
3000 and 6000 atoms with a distribution function given by Eq.
(\ref{eq:rhomathbf-r=rho_0-e-1}).  After that the matrix $M_{jj'}$ is
calculated and processed in order to find the total number of Stokes
photons (\ref{eq:ipp_tt=fr-sum_n-big}). We can then by making a
series of such realizations of the position of the atoms get some
statistics on the inherent noise on the point particle model. In Fig.
\ref{fig:plot-show-comp} we show the result of a numerical calculation
using parameters $\mathcal F=4$ and an optical depth of $d=90$. When
we increase the number of atoms, we decrease the particle density 
in order to keep a fixed Fresnel number and a fixed optical depth. We see from Fig.
\ref{fig:plot-show-comp} that there is some dependence on particle
density, an effect of the fact that the system is a point particle
system and not a continuum, hence we do not expect the analytical
theory developed so far to explain this effect. However as the density desceases
the total radiated power converges.
Finally in Fig. \ref{fig:color-online-here} we compare the total
radiated power in the analytical calculation $P(t)$,
(\ref{eq:i_tt=fr-sum_s}), with the nummerical calculation $P_N(t)$,
(\ref{eq:ipp_tt=fr-sum_n-big}). That the two methods gives very
different results for small times is quite clear since initially the
radiation is dominated by the spontaneous emission, which is not
included in the analytical calculation. At increasing times, which is
the regime where the analytical calculation is supposed to be valid,
the two methods gives quite similar results, and we therefore believe
that the analytical calculation gives an accurate description. 

\begin{figure}
  \centering
  \includegraphics[width=0.48\textwidth]{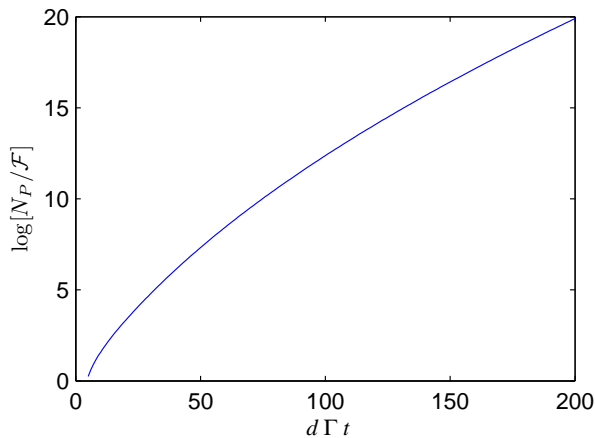}
  \caption{Total number of photons emitted in the
    SRS mode $N_p(t)$ divided by Fresnel number $\mathcal F$ as a
    function of the scaled time $d \Gamma t$. }
  \label{fig:plot-total-number}
\end{figure}

We finally note that for the time-scale used in Fig.
\ref{fig:color-online-here}, the approximation of neglecting depletion
is not completely justified, as the number of emitted photons exceeds
the number of atoms already before the two curves meet. We can examine
the break-down of the no-depletion assumption, by finding the time
$t_c$, at which the number of photons emitted in the superradiating
mode $N_P(t)$ exceeds the number of atoms in the ensemble $N_A$, i.e.
$N_A=N_P(t_c)$, where
\begin{align}\label{eq:n_a=int_0t_c-pt-dt}
  N_P(t)=\int_0^{t} P(t') dt'.
\end{align}
To get an analytical result we will use the approximation $P(t)\approx
P_1(t)$, with $P_1(t)$ given in Eq. (\ref{eq:pt=-fracd-gamma}). After
the integration in Eq. (\ref{eq:n_a=int_0t_c-pt-dt}) we find
\begin{align}
  \frac{2N_P(t)}{\mathcal F} = d\Gamma t \Big[ &I_0^2(\sqrt{d\Gamma t}) -2
  I_1^2(\sqrt{d\Gamma t}) \notag \\ & + I_0(\sqrt{d\Gamma
    t})I_2(\sqrt{d\Gamma t}) \Big]  -
  I_0^2(\sqrt{d\Gamma t}),
\end{align}
where we have used that $d,N_A \gg 1$. In Fig.
\ref{fig:plot-total-number} we plot the function $N_p(T)$. From the
requirement $N_A=N_p$ we find the time $t_c$ for the result shown in
Fig. \ref{fig:color-online-here} to be $\Gamma t_c = 0.54$, where the
radiation is still dominated by spontaneous emission.  If, however we
increase the optical depth we decrease the time at which the
analytical curve $P(t)$, (\ref{eq:i_tt=fr-sum_s}), and the numerical
curve $P_N(t)$, (\ref{eq:ipp_tt=fr-sum_n-big}), agree. For a higher
optical depth $d$ there will thus be a region where the effects
considered here are dominant within the applicability of our theory.
While the limited atom number used here is thus not physically
relevant, the simulation can still be used as a confirmation of the
approximations used in our analytical calculation since both curves
are derived using the same approximation of neglecting depletion of
the atoms.
For the ongoing SRS experiments using
Bose-Einstein condensed atoms e.g. Ref.  \cite{hilliard08} the number
of atoms used in the process is factors of thousands larger than what
we are able to numerically simulate here, and the approximation used
here is  much less severe.

\section{Conclusion}\label{sec:conclusion}

In this paper we have developed a three-dimensional theory for
spontaneous Raman scattering (SRS). The theory applies to an ensemble
of non-moving atoms and is derived by describing the atoms as a
continuous medium. In the theory we neglect the depletion of the
initial atomic state and the theory is therefore mainly applicable to
the onset and build up of SRS. We believe, however, that the theory
still captures the most important effect of the three-dimensional
structure of the problem, since after the onset of SRS the radiation
is dominated by the modes determined by our theory.

The theory is based on a generalization of the one-dimensional theory
in Ref. \cite{raymer-mostow}. In the limit where the Fresnel
$\mathcal{F}$ is very large we find that the one-dimensional
description of Ref. \cite{raymer-mostow} applies to all transverse
modes in agreement with the derivation in Ref. \cite{hammerer}.
Without a detailed investigation of the three-dimensional structure
there is, however, no restriction on the transverse momentum of the
light and a naive application of the theory therefore predicts an
infinite radiated intensity. In our three-dimensional theory the build
up of SRS only happens for small transverse momentum of the light,
limited by the Fresnel number $\mathcal{F}$ of the ensemble. This
automatically limits the emitted radiation such that the theory gives
finite predictions.

In the theory we assume that the dimensions of the ensemble is much
larger than the wavelength of the radiation, and we show that in this
limit the only two parameters describing SRS are the optical depth $d$
and the Fresnel number $F$ of the ensemble. We find that in the limit
$\mathcal{F} \gtrsim 1$ the time scale of SRS is almost exclusively
given by the optical depth $d$ with only a weak dependence on the
Fresnel number $\mathcal{F}$. On the other hand, the total radiated
power for SRS depends strongly on the Fresnel number $\mathcal{F}$.
The total power radiated into modes with a given azimuthal quantum
number $m$ is linearly proportional to $\mathcal{F}$. We find that the
largest contribution to the radiation always comes from the azimuthal
quantum number $m=0$ and that this contribution also have the fastest
growth. With increasing Fresnel number $\mathcal F$ the contribution
from other azimuthal quantum numbers $m\neq 0$ may, however, become
comparable to the $m=0$ contribution. To investigate the validity of
our analytical findings we have compared our analytical results to a
direct numerical solution for a limited number of atoms. The two
approaches are found to be in good agreement.

An interesting question which we have not addressed in detail here
comes from the fact that an ensemble of atom is not given by a
continuous density, but consists of a collection of discrete point
particles. The effect of this is in principle included in our direct
numerical investigations, and may be the reason for the dependence on
the atom number in Fig. \ref{fig:plot-show-comp}. Here the
simulations with the highest density deviate from the results with a
lower density. It would be interesting to investigate such effects
using for instance the methods developed in Ref. \cite{martin-anders}.
Furthermore, the question of the collective emission of radiation from
atomic ensembles is also very interesting from the point of view of
quantum information.  Several important quantum protocols such as
quantum repeaters \cite{DLCZ}, quantum memory \cite{juulsgaard2}, and
quantum teleportation \cite{sherson} are currently being investigated
in atomic ensembles. For a full evaluation of the potential of these
approaches it will by important to have a full understanding of the
effect of the realistic three dimensional structure of the ensembles.
The methods developed in this article may serve as useful starting
point for such investigations.

\acknowledgments 
\noindent 
We thank J.H. Müller, K. Mølmer and J. I. Cirac for usefull
discussions. We acknowledge the financial support of the Future and
Emerging Technologies (FET) programme within the Seventh Framework
Programme for Research of the European Commission, under the FET-Open
grant agreement HIDEAS, number
FP7-ICT-221906.

\appendix

\section{Deriving the first order correction to the matrix
  $M^{kmn}_{k'm'n'}$}
\label{sec:deriving-first-order}

By introducing the dummy variable
$\alpha=2\sigma_{\perp}^2$ in the Gaussian function, the series
expansion of the $x$-integral in Eq. (\ref{eq:m_kmnkmn-=-delta_mm})
may be written as
\begin{align}\label{eq:left.-sum_l=0infty-}
\left.  \sum_{l=0}^{\infty} (-\partial_{\alpha})^l \int_0^{\infty} e^{-\alpha
    x^2} I_m(2\sigma_{\perp}^2\gamma_nx)
  I_m(2\sigma_{\perp}^2\gamma_{n'} x ) \right| _{\alpha=2\sigma_{\perp}^2}
\end{align}
Using the above expansion along with the relation
$I_m(x)=i^{-m}J_m(ix)$ together with the result \cite{integrals}
\begin{align}
    \int_0^{\infty}&rdr e^{-\alpha^2 r^2}J_m(\beta r)J_m(\gamma r) =
  \frac{1}{2\alpha^2} e^{-\frac{\beta^2+\gamma^2}{4\alpha^2}}
  I_m(\frac{\beta \gamma}{2\alpha^2}) \notag \\
  & |\arg[\alpha]|<\frac{\pi}{4}, \Re[m]>-1 ,\beta >0, \gamma >0,
\end{align}
Equation (\ref{eq:left.-sum_l=0infty-}) may be rewritten as
\begin{align}
  \left.\sum_{l=0}^{\infty} \frac{(-\partial_{\alpha})^l}{(-1)^m}
    \int_0^{\infty} e^{-\alpha x^2}J_m(2i\sigma_{\perp}^2\gamma_n
    x)J_m(2i\sigma_{\perp}^2\gamma_{n'}
    x)\right|_{\alpha=2\sigma_{\perp}^2}
\end{align}
From Ref. \cite{integrals} we find the integral to give
\begin{align}
  \left. \sum_{l=0}^{\infty} (-\partial_{\alpha})^l
    \frac{e^{\frac{\sigma_{\perp}^4(\gamma_{n}^2 +
          \gamma_{n'}^2)}{\alpha}}}{2\alpha} I_m\Big( \frac{2
      \sigma_{\perp}^4
      \gamma_n\gamma_{n'}}{\alpha}\Big)\right|_{\alpha=2\sigma_{\perp}^2}.
\end{align}
We see that in terms of an expansion in the variable $ 1/
\sigma_{\perp}^2$ each differentiation will give a factor of
$1/\sigma_{\perp}^2$.  We shall therefore only consider a sum up to the
first order in the differential. To zeroth order the $x$-integral simply
gives
\begin{align}
  \frac{e^{\frac{\sigma_{\perp}^2(\gamma_{n}^2 +
        \gamma_{n'}^2)}{2}}}{4 \sigma_{\perp}^2} I_m\Big(
  \sigma_{\perp}^2 \gamma_n\gamma_{n'}\Big).
\end{align}
To first order we find the $x$-integral to give
 \begin{widetext}
   \begin{align}\label{eq:part-frac-+-1}
   - \partial_{\alpha} 
     \frac{e^{\frac{\sigma_{\perp}^4(\gamma_{n}^2 +
           \gamma_{n'}^2)}{\alpha}}}{2\alpha }& I_m\Big( \frac{2
       \sigma_{\perp}^4
       \gamma_n \gamma_{n'}}{\alpha}\Big)
   \Bigg| _{\alpha =2\sigma_{\perp}^2}   =
     \frac{e^{-\frac{\sigma_{\perp}^2}{2}(\gamma_n^2+\gamma_{n'}^2)}}{8\sigma_{\perp}^4}
     \Bigg[ I_m(\sigma_{\perp}^2\gamma_n\gamma_{n'}) \notag \\ &-
     \frac{\sigma_{\perp}^2}{2}(\gamma_{n}^2+\gamma_{n'}^2)
     I_m(\sigma_{\perp}^2\gamma_n\gamma_{n'})+
     \frac{\sigma_{\perp}^2}{2} \gamma_n\gamma_{n'} \Big(
     I_{m-1}(\sigma_{\perp}^2\gamma_n\gamma_{n'}) +
     I_{m+1}(\sigma_{\perp}^2\gamma_n\gamma_{n'}) \Big) \Bigg].
   \end{align}
 To understand the above expression let us assume a sufficiently large 
$\sigma_{\perp}$ so that the modified Bessel function $I_{m\pm 1}$ can
  be approximated with $I_m$. In this way we get 
    \begin{align}\label{eq:part-frac-+}
      - \partial_{\alpha} \frac{e^{\frac{\sigma_{\perp}^4(\gamma_{n}^2
            + \gamma_{n'}^2)}{\alpha}}}{2\alpha } I_m\Big( \frac{2
        \sigma_{\perp}^4 \gamma_n \gamma_{n'}}{\alpha}\Big) \Bigg|
      _{\alpha =2\sigma_{\perp}^2} =
      \frac{e^{-\frac{\sigma_{\perp}^2}{2}(\gamma_n^2+\gamma_{n'}^2)}}{8\sigma_{\perp}^4}
      I_m(\sigma_{\perp}^2\gamma_n\gamma_{n'}) \Bigg[ 1 -
      \frac{\sigma_{\perp}^2}{2}(\gamma_{n}-\gamma_{n'})^2 \Bigg].
    \end{align}
 \end{widetext}
The above approximation gets worse for increasing values of $m$,
however we argue in Sec. \ref{sec:intensity}, that for a finite width
of the sample, higher order modes in $m$ has less influence. Finally
the exponential function along with the modified Bessel function
express a conservation of transverse momentum given by the variables
$\gamma_n$ since for increasing values of the transverse momentum,
Eq. (\ref{eq:part-frac-+}) can be approximated with 
\begin{align}\label{eq:frace-fracs-gamm}
  \frac{e^{-\frac{\sigma_{\perp}^2}{2}(\gamma_n -
      \gamma_{n'})^2}}{8\sigma_{\perp}^4\sqrt{2\pi
      \gamma_n\gamma_{n'}}} \Bigg[ 1 -
    \frac{\sigma_{\perp}^2}{2}(\gamma_{n}-\gamma_{n'})^2 \Bigg].
\end{align}
We shall then make the approximation
\begin{align}
  1 - \frac{\sigma_{\perp}^2}{2}(\gamma_{n}-\gamma_{n'})^2 \approx
  e^{-\frac{\sigma_{\perp}^2}{2}(\gamma_n - \gamma_{n'})^2},
\end{align}
thus the expression in Eq. (\ref{eq:frace-fracs-gamm}) can to second
order in the difference $\gamma_n-\gamma_{n'}$ be written as
\begin{align}
  \frac{e^{-\sigma_{\perp}^2(\gamma_n -
      \gamma_{n'})^2}}{8\sigma_{\perp}^4\sqrt{2\pi
      \gamma_n\gamma_{n'}}} 
\end{align}
This result is the large size limit, and we therefore conclude that to
give this limit as $\sigma_{\perp} \rightarrow \infty$ the term in
Eq. (\ref{eq:part-frac-+-1}) must be approximated with
\begin{align}
  \sqrt{2} \frac{e^{-\sigma_{\perp}^2(\gamma_n^2 +
      \gamma_{n'}^2)}I_m( 2 \sigma_{\perp}^2 \gamma_n \gamma_{n'} )
  }{8 \sigma_{\perp}^4}.
\end{align}
From this we conclude the result given in Eq.
(\ref{eq:m_kmnkmn-=-delt}).

\section{Commutation relation for $\Lambda^m_{nn'}$ and
  ${\Lambda^1}^m_{nn'}$}
\label{sec:comm-relat-lambd}

Here we show that the two matrices $\Lambda^m_{nn'}$ and
${\Lambda^1}^m_{nn'}$ commute. Since both matrices are symmetric, it
is enough to show that the product
$\sum_{p}\Lambda^m_{np}{\Lambda^1}^m_{pn'}$ is symmetric. Again we
make the continuation $\sum_p \frac{1}{a_c} \rightarrow \int
\frac{d\gamma_p}{\pi}$ for $a_c\rightarrow \infty$. In this way we get
\begin{align}
  \sum_{p}&\Lambda^m_{np}{\Lambda^1}^m_{pn'} =
  \frac{4\sigma_{\perp}^4e^{-\frac{\sigma_{\perp}^2}{2}
      \gamma_n^2-\sigma_{\perp}^2\gamma_{n'}^2}}{a_c^2
      J_{m+1}(X_{mn})J_{m+1}(X_{mn}) } \times \notag \\ & 
    \int d\gamma_p \gamma_p \frac{(-1)^m}{2} e^{-\frac{3
        \sigma_{\perp}^2\gamma_p^2}{2}} J_m( i\sigma_{\perp}^2
    \gamma_n \gamma_p) J_m( 2i\sigma_{\perp}^2
    \gamma_{n'} \gamma_p).
\end{align}
After making the $\gamma_p$-integral we end up with
\begin{align}\label{eq:sum_pl-=-frac4s}
  \sum_{p}&\Lambda^m_{np}{\Lambda^1}^m_{pn'} =
  \frac{4\sigma_{\perp}^4e^{-\frac{\sigma_{\perp}^2}{3}(
      \gamma_n^2+\gamma_{n'}^2)}I_m \Big( \frac{2\sigma_{\perp}^2}{3}
    \gamma_{n} \gamma_{n'}\Big) } {3 a_c^2
    J_{m+1}(X_{mn})J_{m+1}(X_{mn}) }.
\end{align}
Since the matrix Eq. (\ref{eq:sum_pl-=-frac4s}) is symmetric we
conclude that the matrices $\Lambda^m_{nn'}$ and ${\Lambda^1}^m_{nn'}$
commute.

\section{Derivation of Eq. (\ref{eq:int_0-rdr-j_mxrj_mxr}) }
\label{sec:compl-sec.-total}

Here we will show Eq. (\ref{eq:int_0-rdr-j_mxrj_mxr}). Our
starting point is the orthogonality relation given by
\begin{align}\label{eq:int-_0infty-rdr}
  \int _0^{\infty} rdr\: J_m(\gamma_n r)
  J_m(\gamma_{n'}r)=\frac{\delta_{nn'} a_c^2}{2J_{m+1}(X_{mn})^2},
\end{align}
where the variable $\gamma_n=\frac{X_{mn}}{a_c}$ and $X_{mn}$ is the
$n$'th zero of the $m$'th order Bessel function $J_m$. We will assume
that $X_{mn}$ is large, which does not require $\gamma_n$ to be so, since
we can choose the cut-off $a_c$ to be anything. In this way we can write
Eq. (\ref{eq:int-_0infty-rdr}) as
\begin{align}
  \int _0^{\infty} rdr\: J_m(\gamma_n r)
  J_m(\gamma_{n'}r)=\frac{\delta_{nn'} a_c}{\pi \gamma_n}
\end{align}
We will then take the sum over $n$ on both sides and use the standard
continuation $\sum_n \frac{1}{a_c}\rightarrow \int \frac{d
  \gamma_n}{\pi}$ so that
\begin{align}
  \int d\gamma_n \gamma_n\int rdr J_m(\gamma_n r)
  J_m(\gamma_{n'}r)=1.
\end{align}
Since $\gamma_n$ is now a continuous variable, we conclude that the
measure of the distribution
\begin{align}
  f(x,x')=x\int rdr J_m(x r) J_m(x' r),
\end{align}
where $x,x'$ is some real and positive number is unity.  The next step
is to show that for $x\neq x'$ the function $f(x,x')$ vanish.  This
follows when choosing a zero point $X_{mn}$ and a cut-off $a_c$ such
that say $x=\gamma_n$.  This does not necessarily mean that $x'$ has a
similar representation with the chosen cut-off. On the other
hand this is not necessary as one may show, see e.g.  \cite{jackson},
that
\begin{align}
  (\gamma_n^2-{x'}^2 )\int_0^{a_c} rdr J_m(\gamma_nr)J_m(x'r)=0.
\end{align}
from here we conclude that when $\gamma_n$ and $x'$ are different the
function $f(\gamma_n,x')$ vanish. This concludes the derivation of
Eq. (\ref{eq:int_0-rdr-j_mxrj_mxr}).

\section{The Sum rule}\label{sec:sum-rule}

Here we derive the sum rule Eq. (\ref{eq:frac2kl-hbar-epsil}) used in
Sec. \ref{sec:total-coher-radi}.  The starting point is the total
radiated intensity of Stokes-photons
\begin{align}\label{eq:oint_s-big-mathbf}
  \oint_S \Big\{ \mathbf D^- \times \big( \nabla \times \mathbf A^+
  \big) - \big( \nabla \times \mathbf A^- \big) \times \mathbf D^+ \Big\},
\end{align}
where $S$ is a sphere surrounding the atoms.
Using the Divergence theorem as well as the Maxwell equations, the total
radiated intensity can be written as
\begin{align}
  -\mu_0\epsilon_0 \int_V d^3r \frac{\partial (\mathcal H_F + \mathcal
    H_I)}{\partial t} -\mu_0 \int_V d^3r
  G[\mathbf P,\mathbf D] \label{eq:mu_0-int_v-d3r} \intertext{where}
G[\mathbf P,\mathbf D] = \frac{\partial \mathbf P^-}{\partial
    t} \cdot \mathbf D^+   +  \mathbf D^- \cdot \frac{\partial \mathbf
    P^+}{\partial t}.
\end{align}
To lowest order in $1/\stokes$, Eq. (\ref{eq:mu_0-int_v-d3r}) reduce to 
\begin{align}
  \mu_0\stokes\hbar\epsilon_0\Big[& \sum_j \Gamma \hat b_j(t) \hat
  b_j^{\dagger}(t) \notag \\ & + \sum_{j\neq j'} \left\{ 
    \hat b_j(t) M_{jj'} \hat b_{j'}^{\dagger}(t) + H.c. \right\} \Big],
\end{align}
where we have used Eqs. (\ref{eq:mathbf-p+mathbf-r}), (\ref{eq:m_jj-=-frac})
and (\ref{eq:c2-mathbf-e}).  When measuring the intensity infinitely far
away from the atomic ensemble, the expression in Eq.
(\ref{eq:oint_s-big-mathbf}) reduce to the electric field squared
times $2\mu_0c$, thus the normalized sum-rule reads
\begin{align}
  \frac{2}{\kstokes \hbar \epsilon_0 } \int d\Omega \mathbf D^-
  \cdot&\; \mathbf D^+ = \sum_j \Gamma \hat b_j(t) \hat
  b_j^{\dagger}(t) \notag \\ & + \sum_{j\neq j'} \left\{ 
    \hat b_j(t) M_{jj'} \hat b_{j'}^{\dagger}(t) + H.c. \right\}.
\end{align}

\end{document}